# Experimental and theoretical study of structural properties and phase transitions in $YAsO_4$ and $YCrO_4$

D. Errandonea[1], R. Kumar[2], J. López-Solano[3], P. Rodríguez-Hernández[3], A. Muñoz[3], M. G. Rabie[4], and R. Sáez Puche[4]

[1]*MALTA Consolider Team, Departamento de Física Aplicada-ICMUV, Universitat de Valencia, Edificio de Investigación, c/Dr. Moliner 50, Burjassot, 46100 Valencia, Spain*

[2]*High Pressure Science and Engineering Center, Department of Physics and Astronomy, University of Nevada Las Vegas, 4505 Maryland Parkway, Las Vegas, Nevada 89154-4002, USA*

[3]*MALTA Consolider Team, Departamento de Física Fundamental II,* and Instituto de Materiales y Nanotecnología, *Universidad de La Laguna, La Laguna, Tenerife, Spain*

[4]*Departamento de Química Inorgánica I, Facultad de Ciencias Químicas, Universidad Complutense de Madrid, 28040 Madrid, Spain*

**Abstract:** We have performed experimental and theoretical studies of the structural stability of $YAsO_4$ and $YCrO_4$ at high pressures. X-ray diffraction experiments together with *ab initio* total-energy and lattice-dynamics calculations have allowed us to completely characterize a pressure-induced structural phase transition from the zircon to the scheelite structure in both compounds. Furthermore, total-energy calculations have been performed to check the relative stabilities of different candidate structures at different pressures and allow us to propose for $YAsO_4$ the zircon → scheelite → $SrUO_4$-type sequence of structures. In this sequence, sixfold arsenic coordination is attained for the $SrUO_4$-type structure above 32 GPa. The whole sequence of transitions is discussed in comparison with $YVO_4$, $YPO_4$, $YNbO_4$, $YMoO_4$, and $YTaO_4$. Also a comparative discussion of lattice-dynamics properties is presented. The band-gap for $YAsO_4$ and $YCrO_4$ and band structure for $YAsO_4$ are also reported. Finally, the room-temperature equation of state of different compounds is also obtained.

PACS number(s): 62.50.-p, 61.05.cp, 61.50.Ks

## I. Introduction

Yttrium arsenate ($YAsO_4$) and yttrium chromate ($YCrO_4$) belong to the family of oxides of general formula $RXO_4$, where R is a trivalent metal and X = P, As, Cr, or V. They mostly crystallize in two structural types, either zircon or monazite, depending on the sizes of the trivalent metal and the X element [1]. In particular, $YAsO_4$ (the mineral chernovite) and $YCrO_4$ have an isomorphic crystal structure to zircon. The structure belongs to the tetragonal space group (S.G.) $I4_1/amd$ and has four formula units per unit-cell (Z = 4). The zircon structure (see Fig. 1) is prototypic of many ternary oxides [1]. The R atoms are eight coordinated to oxygen (O) atoms forming bisdisphenoids (dodecahedra) and the X atoms coordinated to four O atoms forming tetrahedra [2]. These two sorts of polyhedral units are edge-connected, and alternate in rods along the *c*-axis. Zircon-structured oxides are important to a variety of geophysical and geochemical applications as they are common accessory minerals in a large number of rock types in the Earth's upper mantle [3]. Additionally, they have a wide potential for their application in optoelectronics due to their optical and luminescent properties [4]. The structural stability and thermodynamic properties of many of these oxides have been systematically investigated. In particular, high-pressure (HP) studies have been performed on them in order to understand their mechanical properties and HP structural behavior [5]. In these studies the occurrence of phase transitions has been observed in silicates [6], germanates [7], chromates [8, 9], phosphates [10, 11], and vanadates [12, 13]. In $YVO_4$ (the mineral wakefieldite), *in-situ* HP x-ray diffraction (XRD) studies detected a non-reversible phase transition at 8.5 GPa [13]. The transition is from the zircon structure to another tetragonal structure which belongs to space group $I4_1/a$. The HP structure is isomorphic to that of the mineral scheelite ($CaWO_4$) [2]. In $YCrO_4$, a similar transition was found at 3 GPa [8]. In contrast, $YPO_4$ (the mineral xenotime) is found to remain stable in the zircon-type structure up to pressures higher than 16 GPa [10, 11]. In addition, in this case the transformation is not from zircon to scheelite, but to a monoclinic monazite-type structure (S.G. $P2_1/n$). Upon further compression a transition from monazite to scheelite is found at 32 GPa [11]. Despite the phase transitions in $YVO_4$, $YCrO_4$, and $YPO_4$, the effects of pressure on the crystalline structure of $YAsO_4$ have not been studied before our work here.

In this work, to gain further understanding of the structural properties of zircon-type oxides, HP XRD experiments on zircon- type $YAsO_4$ and $YCrO_4$ up to 20 GPa are reported. A*b initio* total-energy calculations are also reported for $YAsO_4$. In both compounds we detected the occurrence of a zircon-scheelite transition. Our calculations predict an additional phase transition for $YAsO_4$ at 32 GPa to an structure with six coordinated arsenic. The equations of state (EOS) for the different structures are presented too. Furthermore, the EOS for related compounds $YMoO_4$, $YNbO_4$, and $YTaO_4$ are calculated. All these results are compared with those previously found in $YVO_4$, $YCrO_4$, and $YPO_4$. Finally, Raman spectroscopy measurements and lattice-dynamics calculations are also reported for $YAsO_4$ and its electronic structure characterized.

## II. Experimental details

HP XRD experiments in $YAsO_4$ were performed on powder samples obtained from a natural crystal provided by Excalibur Mineral Company and collected at the Sierra County, New Mexico, USA. Electron microprobe analysis was performed to determine the impurities present in the natural crystal. The main impurities were La (0.8%) and P (0.6%). Ambient conditions XRD measurements, using Cu $K_\alpha$ monochromatic radiation ($\lambda$ = 1.5406 Å), confirmed that $YAsO_4$ has a zircon-type structure. The obtained structural parameters are summarized in Table I. They agree with the parameters reported in the literature [14, 15]. In particular, $a$ = 7.070(6) Å is closer to the value reported by Strada *et al.* [14] and $c$ = 6.214(4) Å is close to the value reported by Goldin *et al.* [15]. Natural $YAsO_4$ was also characterized by Raman spectroscopy. To perform these measurements we used the 514.5 nm excitation line of an Ar-ion laser operated at a power of 70 mW. We detected nine out of the twelve Raman-active modes of the zircon structure [16]. Their frequencies are summarized in Table II. They agree well with those measured in pure synthetic $YAsO_4$ [17]. The agreement observed in both Raman and XRD measurements with previous studies indicate that as a first approximation we can neglect the effects of impurities on the structural properties of $YAsO_4$.

HP XRD experiments in $YCrO_4$ were performed on synthetic powder samples. Powder zircon-type $YCrO_4$ was prepared by heating stoichiometric amounts of $Cr(NO_3)_3.9H_2O$ (99%) and $Y(NO_3)_3.6H_2O$ (99.9%) in a oxygen flow which is required to stabilize the unusual oxidation state $Cr^{5+}$. A thermal treatment has been carried out as follows: 30 min at 433 K, 30 min at 473 K and 10 h at 853 K. Afterwards, the sample was cooled down inside the furnace in the oxygen flow for 6 h. The synthesized samples were characterized by XRD at ambient conditions using Cu $K_\alpha$ monochromatic radiation. It was confirmed that $YCrO_4$ has the zircon-type structure with no indication of additional phases. The unit-cell parameters were $a$ = 7.089(5) Å and $c$ = 6.281(3) Å. The obtained atomic positions are summarized in Table I. All the structural parameters agree well with those previously reported [8]. Raman measurements also confirmed the zircon structure of $YCrO_4$. The detected Raman-active modes are summarized in Table II and compared with earlier data [18].

HP angle-dispersive XRD measurements were carried out at room temperature (RT) in a 300-μm culet diamond-anvil cell (DAC). The powder samples were loaded together with a ruby chip into a 100-μm-diameter hole drilled on a 200-μm-thick rhenium gasket pre-indented to 35 μm. A 16:3:1 methanol-ethanol-water mixture was used as pressure-transmitting medium [19, 20] and the pressure was determined using the ruby fluorescence technique [21]. The experiments were performed at the 16-IDB beamline of the HPCAT facility at the Advanced Photon Source (APS) using monochromatic radiation ($\lambda$ = 0.3931 Å). The x-ray beam was focused down to 15 x 10 $\mu m^2$ by using Kickpatrick–Baez mirrors and spatially collimated with a 30 μm molybdenum cleanup pinhole. Diffraction images were recorded with a Mar345 image plate detector, located 350 mm away from the sample, and were integrated and corrected for distortions using FIT2D [22]. Typical diffraction patterns were collected with 20 s exposures. Two independent runs were performed for each of the studied oxides. The indexing, structure solution, and refinements were performed using the FULLPROOF [23] and POWDERCELL [24] program packages. Finally, the electronic band-gap of $YAsO_4$ and $YCrO_4$ is estimated at ambient conditions from reflectivity and optical-absorption measurements [25].

## III. Overview of the calculations

The total-energy and lattice-dynamics calculations presented in this work have been performed with the VASP *ab initio* code [26, 27], which solves the many-body problem within the density-functional formalism using pseudopotentials and a plane-waves basis set. In these calculations, pseudopotentials generated with the projector-augmented wave method [28] within the Perdew-Burke-Ernzerhof (PBE) [29] approximation for the exchange-correlation energy were used and a plane-wave energy cutoff of 520 eV was employed. Together with dense Monkhorst-Pack [30] grids appropriate for each structure, an energy convergence of ~1 meV per formula unit (pfu) was obtained. At selected volumes, the atomic and lattice degrees of freedom of each structure were optimized until the atomic forces were lower than ~0.001 eV/Å, and the deviation of the stress tensor from the diagonal hydrostatic form was less than 1 kbar (0.1 GPa). With the values of energy and pressure for a set of volumes, the enthalpy at each pressure was calculated to determine the most stable phase among the ones considered. Temperature effects were not taken into account.

Raman and infrared modes have been calculated within the direct force constant method [31]. The PHONON code [32] was used to generate a minimal set of symmetry non-equivalent atomic displacements from which the full dynamical matrix was constructed. Diagonalization of the dynamical matrix then provided the frequencies and polarization vectors of the vibrational modes at the zone center.

## IV. Results

### A. Structural studies of $YAsO_4$

Fig. 2 shows a selection of diffraction patterns of $YAsO_4$ measured at different pressures. A closely similar pressure evolution was observed for the XRD patterns of $YCrO_4$. Considering a zircon-type structure, all the diffraction patterns observed between ambient pressure and 7 GPa can be well indexed. However, at 8 GPa we observed the appearance of weak peaks in addition to the peaks of zircon. One example of these new reflections is the one at 2θ = 7.5º depicted by a cross in the figure. The intensities of the emergent peaks gradually increase from 8 GPa to 12 GPa. At the same time the peaks of zircon gradually lose intensity and fully disappear at 14 GPa. Typical weak peaks of zircon phase are indicated with an asterisk in the figure. These results indicate that a phase-transition takes place in $YAsO_4$. The onset of the transition is at 8 GPa, but the transformation is not fully completed up to a pressure higher than 12 GPa. Within this range of pressure, the zircon phase coexists with the high-pressure phase. The transition is gradual as shown by the continuous change of peak intensities. From 14 to 18.9 GPa there are no additional changes in the diffraction patterns of $YAsO_4$ indicating that the HP phase is stable up to this pressure. On pressure release, the phase transition is non-reversible as illustrated by the diffraction pattern measured at ambient pressure and shown in the upper trace of Fig. 2.

Compounds isomorphic to $YAsO_4$ also undergo pressure-induced phase transitions. In particular, $YVO_4$ transforms from zircon to scheelite [13] and $YPO_4$ transforms from zircon to monazite [10, 11]. Consequently, in order to characterize the crystalline structure of the HP phase of $YAsO_4$ we have considered these two structures as candidates. After a deep examination of the diffraction patterns assigned to the HP phase, we found that they can be well indexed considering a scheelite-type structure (S.G $I4_1/a$, Z = 4). At 14 GPa we obtain for it the following unit-cell parameters $a$ = 4.902(5) Å and c = 10.821(9) Å. The structural parameters obtained for the scheelite structure at ambient pressure, after pressure release, are shown in Table I. The transition from zircon-to-scheelite is of first order and involves a volume collapse of 10.5%.

From our experiments we extracted the pressure evolution of the unit-cell parameters for both phases. The results are summarized in Fig. 3. As can be seen in the figure, the compression of the zircon-type structure is non-isotropic, being the *c*-axis the less compressible axis. As a consequence of this, in the zircon phase the axial ratio (*c/a*) gradually increases from 0.879 at ambient pressure to 0.889 at 12 GPa. A similar behavior was found for $YVO_4$ and $YPO_4$ [11, 13]. Regarding the unit-cell parameters of the scheelite-type structure, we found that compression is also anisotropic, being the *c*-axis the most compressible axis (as it happens in many other scheelites) [33, 34]. In particular, *c/a* decreases from 2.240 at ambient pressure to 2.181 at 18.9 GPa.

From the pressure dependence of the lattice parameters, the unit-cell volumes of different phases of $YAsO_4$ as a function of pressure were also calculated. The results are summarized in Fig. 4. We have analyzed the volume changes using a third-order Birch-Murnaghan EOS [35]. The EOS parameters for phase I are: $V_0$ = 310.6(4) Å$^3$, $B_0$ = 142(3) GPa, and $B_0$'= 4.1(7), being these parameters the zero-pressure volume, bulk modulus, and its pressure derivative, respectively. The implied value for the second derivative of $B_0$ ($B_0''$) is -2.82 $10^{-2}$ GPa$^{-1}$. The bulk modulus of zircon-type $YAsO_4$ is comparable with that of zircon-type $YPO_4$ and $YVO_4$ [11, 13] and similar to the value calculated for $YAsO_4$ [36]. The EOS parameters for the HP phase are: $V_0$ = 277.4(5) Å$^3$, $B_0$ = 149(6) GPa, and $B_0$'= 4.3(9). The implied $B_0''$ is -2.87 $10^{-2}$ GPa$^{-1}$. The EOS fits for both phases are shown as solid lines in Fig. 4. The bulk modulus of the scheelite phase is similar to that reported for the scheelite-type phase in $YVO_4$ [13]. Empirical models have been developed for predicting the bulk moduli of zircon- and scheelite-structured $ABO_4$ compounds [37]. In particular, the bulk modulus of $YAsO_4$ can be estimated from the charge density of the $YO_8$ polyhedra using the relation $B_0 = 610 Z_i / d^3$, where $Z_i$ is the cationic formal charge of yttrium, *d* is the mean Y–O distance at ambient pressure (in Å), and $B_0$ is given in GPa [37]. Applying this relation a bulk modulus of 139(25) GPa is estimated for the low-pressure phase of $YAsO_4$ and a bulk modulus of 148(29) GPa is estimated for the scheelite-type phase. These estimations reasonably agree with the values obtained from our experiments and indicate that the scheelite-type phase is less compressible than the zircon-type phase. This fact is consistent with the increase of density that occurs together with the transition.

## B. Structural studies of $YCrO_4$

The present results of high-pressure structural studies for $YCrO_4$ qualitatively agree with our own results on $YAsO_4$. In particular, in our experiments the peaks identified with the HP phase were found at 4.2 GPa and the low- and high-pressure phases are found to coexist up to 7 GPa. A pure diffraction pattern of the HP phase is only observed at 8.2 GPa. These results agree well with the conclusions obtained from HP Raman experiments [18]. As illustrated in Fig. 2 for $YAsO_4$, the transition in $YCrO_4$ is sluggish, changing continuously the intensities associated to the Bragg peaks of the low- and high-pressure phases. The HP phase remains stable up to 18.9 GPa and the phase transition is non-reversible. As proposed by Long *et al.* [8, 18], and in agreement with the behavior of $YAsO_4$ and $YVO_4$, we assigned a scheelite-type structure to the HP phase, being the unit-cell parameters at 8.2 GPa: $a$ = 4.941(5) Å and $c$ = 10.996(9) Å. This implies the existence of a large volume collapse of about 11.5 % at the zircon-to-scheelite transition. The structural parameters obtained for the scheelite structure at ambient pressure are shown in Table I. They agree with the results obtained from quenched sample by Long *et al.* [8].

We would like to comment now on the non-reversibility of the zircon-scheelite transition. In both $YAsO_4$ and $YCrO_4$ we found that the scheelite structure remains metastable after decompression. The same phenomenon was observed in many zircon-structured oxides [6 -12]. This is a notable aspect of the phase transformation from zircon to scheelite as it is also the large volume collapse that takes place at the transition. In addition, the irreversibility of the transition is extreme as observed for $ZrSiO_4$ [6]. In this compound, to return the scheelite structure to the zircon structure, at ambient pressure, temperatures larger than 1273 K are required [6]. Therefore, a large kinetic barrier should be involved with this transformation, which is consistent with the reconstructive mechanism of this transition and its first-order character.

The pressure evolution of the unit-cell parameters of both $YCrO_4$ phases are summarized in Fig. 3. As in the case of $YAsO_4$, $YVO_4$ [13], and $YPO_4$ [11] the compression of $YCrO_4$ is anisotropic, the $a$-axis being more compressible in the zircon phase and the *c*-axis being more compressible in the scheelite phase. As a consequence of it, in the zircon-type phase, the *c*/*a* ratio increases from 0.886 at ambient pressure to 0.893 at 7 GPa. In the scheelite-type phase it decreases from 2.250 at ambient pressure to 2.193 at 18.9 GPa. From the pressure dependence of the unit-cell parameters, the volume of the different phases of $YCrO_4$ as a function of pressure is calculated. A summary can be found in Fig. 4. We have analyzed the volume changes using a third-order Birch-Murnaghan EOS [35] and the obtained EOS parameters for phase I are: $V_0$ = 315.6(4) Å$^3$, $B_0$ = 136(4) GPa, and $B_0$' = 4.7(9); implied $B_0''$ = -3.73 $10^{-2}$ $GPa^{-1}$. The bulk modulus is comparable with those of $YAsO_4$, $YVO_4$, and $YPO_4$ and the value obtained using the phenomenological model of Ref. [37] (139 GPa) but a 15 % larger than that obtained from Li *et al. ab initio* calculations (121.6 GPa) [38]. The obtained EOS parameters for the HP phase are: $V_0$ =281.9(3) Å$^3$, $B_0$=151(4) GPa, and $B_0$' = 4.2(6); implied $B_0''$ = -2.73 $10^{-2}$ $GPa^{-1}$. These values are also comparable with the parameters obtained in the scheelite phases of $YAsO_4$, $YVO_4$ [13], and $YPO_4$. The EOS fits for both phases are shown as solid lines in Fig. 4. A comparison of different values of the bulk moduli is shown in Table III. Note that again the scheelite phase is less compressible than the zircon phase.

### C. Electronic properties

Optical-absorption measurements were performed at ambient conditions for $YAsO_4$. We found that the absorption coefficient exhibit a steep absorption, characteristic of a direct band-gap, plus a low-energy absorption band which overlaps partially with the fundamental absorption. This low-energy absorption band has been previously observed in related tungstates and seems to be caused by the presence of defects or impurities [25]. From the absorption measurements we determined the band-gap energy to be 4.5 eV. This result was confirmed by reflectance measurements performed in the same sample. In the case of $YCrO_4$, since we used powder samples, only reflectivity measurements were performed. For this compound we determined $E_g$ = 3.2 eV. Unfortunately, there is no data available in the literature on the band-gap of zircon-type arsenates and chromates to compare with. Therefore, we will discuss our results in the next section in comparison with band-structure calculations.

**D. *Ab initio* results**

For our theoretical study of the structural stability of $YAsO_4$ at high pressures, we have taken into consideration previous results obtained in the $RXO_4$ family of compounds and the packing-efficiency criterion proposed by Bastide [5]. We have studied the relative stability of the zircon (S.G. No. 141, $I4_1/amd$, $Z$ = 4), scheelite (S.G. No. 88, $I4_1/a$, $Z$ = 4), and monazite (S.G. No. 14, $P2_1/n$, $Z$ = 4) structures, using the calculation method outlined in section III. Other structures considered in the calculations are: M-fergusonite (S.G. No. 15, $I2/a$, $Z$ = 4), $LaTaO_4$–type (S.G. No. 14, $P2_1/c$, $Z$ = 4), $BaMnF_4$-type (S.G. No. 36, $A2_1am$, $Z$ = 4), $BaWO_4$-II-type (S.G. No. 14, $P2_1/n$, $Z$=8), $CuTeO_4$-type (S.G. No. 14, $P2_1/n$, $Z$ = 4), iwashiroite ($YNbO_4$-type, S.G. No. 13, $P2/$a, $Z$ = 2), and $SrUO_4$-type (S.G. No. 57, $Pbcm$, $Z$ = 4). Fig. 5 shows the energy as a function of volume curves for the calculated structures. In Fig. 6 we illustrate the enthalpy-pressure curve computed for $YAsO_4$. From these plots the relative stability and coexistence pressures of the different phases can be extracted. We found, in agreement with the experiments, that zircon is the stable structure of $YAsO_4$ at ambient pressure. The unit-cell parameters for this phase are $a$ = 7.171 Å and $c$ = 6.371 Å and the atomic coordinates of the O atoms are (0, 0.43101, 0.19790). The agreement with the experimental results is comparable with that obtained in similar oxides. In particular the small overestimation of the unit-cell parameters is comparable with typical differences between GGA calculations and experiments. A fit of the energy versus volume curve with a Birch-Murnaghan fourth-order EOS [35] gives the following EOS parameters for the zircon structure $V_0$ = 328.23 Å$^3$, $B_0$ = 115.60 GPa, $B_0'$= 6.22, and $B_0''$ = -9.55 $10^{-2}$ GPa$^{-1}$. Present and previous experimental and theoretical results are compared in Table III. Calculations underestimate the bulk modulus at ambient pressure by 15% as happens in $YCrO_4$ and $YVO_4$.

As shown in the enthalpy as a function of pressure curves of Fig. 6, our calculations show that the scheelite phase becomes more stable than zircon at 4 GPa. The monazite structure is unstable in all the pressure range covered by our studies. The observed transition is a first-order transition in which the volume change is 11.2%. This is in agreement with the experimental results. The difference in the transition pressure may be caused by the presence of kinetic-energy barriers or temperature effects as observed in other $RXO_4$ oxides [42]. The unit-cell parameters for this HP phase at ambient pressure are $a$ = 5.053 Å and $c$ = 11.357 Å and the atomic coordinates of the O atoms are (0, 0.10113, 0.04602). The agreement with the experimental results is comparable with that obtained for the zircon phase. A fit of the energy versus volume curve with a Birch-Murnaghan fourth-order EOS [35] gives the following EOS parameters for the scheelite structure $V_0$ = 291.21 Å$^3$, $B_0$ = 135.22 GPa, $B_0'$= 5.61, and $B_0''$ = -5.98 $10^{-2}$ GPa$^{-1}$. A comparison among present and previous experimental and theoretical results is made in Table III.

In Figs. 5 and 6, it can be also seen that a second phase transition is predicted to take place in $YAsO_4$. In this case the transition is from scheelite to an orthorhombic structure isomorphic to that of $SrUO_4$. The transition takes place at 32 GPa, with 7.5% change in the volume. It is interesting to note, that in contrast with $YVO_4$, we found that the M-fergusonite structure does not become the most stable phase in any pressure range. Indeed, according with the calculations, this structure (a monoclinic distortion of scheelite) always reduces to scheelite. The calculated structural parameters for the structure $SrUO_4$-type at 34.5 GPa are summarized in Table IV. The parameters of a fourth-order Birch-Murnaghan EOS fitted to the energy-volume data of the $SrUO_4$-type phase are $V_0$ = 263.43 Å$^3$, $B_0$ = 171.9 GPa, $B'_0$ = 4.3, and $B''_0$ = -0.86 $10^{-4}$ GPa$^{-1}$. The $SrUO_4$-type structure is orthorhombic and belongs to the $n$ = 1 Dion-Jacobson structures [43]. The Sr atoms (Y in our case) atoms are sandwiched between infinite $UO_6$ ($AsO_6$) sheets [44]. It can be seen as a structure made of perovskite-like slabs separated by Y cations. The appearance of this structure at HP implies an increase of the coordination of both cations in constrast with the zircon-to-scheelite transition where no coordination change occurs. It also implies an increase of the packing efficiency in contrast with zircon and scheelite as can be seen in Fig. 1. At pressures where it is stable, the structure features Y in 9+1 coordination and $AsO_6$ units in octahedral coordination, with two As-O bond lengths of 1.72 Å, two of 1.83 Å, and another two of 1.88 Å, the next neighbors being at distances above 3 Å. The $AsO_6$ octahedron is thus distorted with a short axial distance of 3.44 Å and two long equatorial distances with an average length of 3.71 Å. It should be noted that the same structure was found to be stable as a post-scheelite phase in $TbPO_4$ [45].

We have also calculated the Raman active and infrared (IR) active phonons for the zircon phase of $YAsO_4$. Group theory predicts the following vibrational representation for zircon at the Γ point: $\Gamma_{zircon} = (2A_{1g} + B_{1u}) + (B_{2g} + A_{1u}) + (A_{2g} + 2B_{2u}) + (4B_{1g} + 4A_{2u}) + (5E_g + 5E_u)$. Among these 26 modes 12 are Raman active $\Gamma_{zircon}= 2A_{1g} + 4B_{1g} + B_{2g} + 5E_g$ and 7 are IR active $3A_{2u} + 4E_u$. The rest of the modes are acoustic ($A_{2u} + E_u$) or silent modes ($B_{1u} + A_{2g} + A_{1u} + 2B_{2u}$). In Table II we present the calculated Raman-active phonons together with calculated pressure coefficients. In Table V we present the same information for the IR-active and silent phonons. The Grüneisen parameter was calculated considering the experimental bulk modulus (142 GPa). Figs. 7 and 8 show the pressure dependence of IR and Raman phonons. As can be seen in the Tables, the Raman and IR phonon frequencies agree within a 10% with the experimental results at ambient pressure. Unfortunately, there are not HP experiments to compare with. However, discrepancies in the pressure coefficients are usually of the same order than in the phonon frequencies [16, 45]. So our calculations can be trusted as a good estimation of the evolution of the phonons under compression. There are several features to consider from these results. First, the pressure evolution of different Raman modes is similar to that found in $YVO_4$ [41]. In particular, the high-frequency modes have a larger pressure coefficient than the rest of the modes. The IR modes behave qualitatively in the same way. Other interesting fact is that the $E_g$ mode with a frequency of 234 $cm^{-1}$ has a larger pressure coefficient than the other low-frequency modes. Indeed, its Grüneisen parameter is the largest one and due to its large pressure coefficient, there is a phonon crossing with a $B_{2g}$ mode (255 $cm^{-1}$ at ambient pressure) near 2 GPa. We would also like to note that we found three Raman and two silent modes that have negative pressure coefficients. This soft-mode behavior has been observed before in other zircon-structured oxides. It seems to be related to mechanical instabilities induced by pressure in the zircon structure.

To conclude with the report of the theoretical results we would like to present the electronic band structure of zircon-type $YAsO_4$. Band dispersions are plotted along different symmetry directions within the Brillouin zone in Fig. 9. There it can be seen that the valence-band maxima and conduction-band minima are located at the Γ point, so that $YAsO_4$ is a direct-gap material. The calculated value for the band-gap energy ($E_{gap}$) is 3.6 eV which is smaller than the value we obtained from our optical measurements ($E_{gap}$ = 4.5 eV). In the case of $YCrO_4$, previous calculations also gave a smaller $E_{gap}$ than the value we obtained from the experiments ($E_{gap}$ = 3.2 eV). These differences are not surprising since the values of the band gaps calculated within density-functional theory are known to be underestimated [46]. According with the results, here and previously reported [47 – 49], the ordering of the band gaps in the studied oxides is given by $YCrO_4$ (3.2 eV) < $YVO_4$ (3.8 eV) [47] < $YAsO_4$ (4.5 eV) < $YPO_4$ (6.1-9.2 eV) [48, 49]. The only difference between the different oxides is the electronic configuration of the pentavalent atom, therefore it should contribute to dominating states near the Fermi level. This idea is supported by the comparison of the partial density of states of $YVO_4$ and $YAsO_4$. In Fig. 10 we show the partial density of states of $YAsO_4$. There it can be seen, than in contrast with $YVO_4$, where V *3d*-states and O *2p*-states [47] mainly contribute to the bottom part of the conduction band and the upper part of the valence band, in $YAsO_4$ these portions of the bands are dominated by Y *4d*-states and O *2p*-states. Based upon this comparison we expect Y *4d*-states and O *2p*-states to dominate the band structure near the band edges in $YPO_4$, but Cr *3d*-states and O *2p*-states to do it in $YCrO_4$. On the basis of this conclusion, it can be suggested that cation substitution can be used as an efficient tool for band-gap manipulation in $YXO_4$ phosphors. Probably, also a mixed composition of the Y(As,Cr,P,V)$O_4$ system could be used for tuning the band gap of $YXO_4$ phosphors.

## V. Discussion

### A. Structural stability and local compressibility

Here we will compare results obtained in $YAsO_4$, $YCrO_4$ and related compounds. Fig. 11 summarizes the structural sequence of $YAsO_4$, $YCrO_4$, $YPO_4$, and $YVO_4$. The compounds have been ordered in the vertical axis by increasing the ionic radii of the pentavalent atom from the bottom to the top. The four compounds crystallize in the zircon structure. Most of them transform to scheelite under compression, but $YPO_4$ does it through the intermediate monazite structure. This behavior resembles that found in orthophosphates [11]. In these compounds when the radii of the trivalent cation is small (e.g. Sc) a zircon-to-scheelite transition occurs upon compression. However, when the size of the trivalent cation is enlarged, the zircon-monazite-scheelite sequence is observed. Indeed those compounds where the R cation is larger than Tb crystallize in the monazite structure at ambient conditions. The different structural sequence of zircon-type $RXO_4$ oxides can be understood using crystal chemistry arguments [5]. It has been discussed by Errandonea and Manjon [5] that the structural behavior of $RXO_4$ compounds can be rationalized as a function of the ratio of R/X cation sizes. In fact, trends in HP phase transformations can be predicted for a given compound considering the structures of compounds with larger R and X cations [5]. In our case, this systematic is consistent with the fact that the zircon-monazite-scheelite sequence is found for $YPO_4$, which has an R/X ratio more similar to monazite ($CePO_4$) than to scheelite ($CaWO_4$). It also explains the zircon-scheelite sequence found in $YAsO_4$, $YCrO_4$, and $YVO_4$, which are close to the phase boundary between zircon and scheelite in the phase diagram proposed in Ref. 5. The zircon-scheelite-fergusonite sequence (with the following group-subgroup relationship $I4_1/amd \subset I4_1/a \subset I2/a$) can be also accounted by the described systematic. Fig. 11 can be seen as a simplified version of the structural phase diagram for R = yttrium. Indeed it can be seen that the studied oxides take under pressure the scheelite-type structure of $YMoO_4$ [50], having Mo a larger atomic radii than V. Subsequently, $YVO_4$ takes the M-fergusonite (also known as β-fergusonite) structure of $YNbO_4$ and $YTaO_4$ [41], having Nb and Ta a much larger radii than Mo. Fig. 11 also explain the fact that at high-temperature (volume expansion) monoclinic $YNbO_4$ transforms into a tetragonal scheelite-type structure [51].

Obviously, the arguments presented above only give part of the story since aspects like regularly and distortion of polyhedral must be also considered, but it can be used to predict additional phase transitions. In the case of $YAsO_4$ calculations indicate that the post-scheelite phase has a $SrUO_4$-type structure (not appearing fergusonite in between). The appearance of this phase at much higher pressure is consistent with our crystal-chemistry arguments. Uranium has a larger radius than any of the compounds here studied and Sr has a larger radius than Y. This result is then suggesting than the $SrUO_4$-type is a strong candidate as post-scheelite phase for $YPO_4$. Indeed, the zircon-monazite-scheelite-$SrUO_4$ sequence has been already theoretically predicted for $TbPO_4$ [45]. Fig. 11 also suggests that $YCrO_4$, in contrast with $YAsO_4$, should follow the zircon-scheelite-fergusonite sequence like $YVO_4$. In fact, the phonon softening of a $B_g$ Raman mode, typical of the scheelite-fergusonite transition, has been observed under compression in $YCrO_4$, supporting this hypothesis [52]. It will be also reasonable to predict that fergusonite will be the HP phase of $YMoO_4$ and that high coordination phases will appear under compression in $YMoO_4$, $YNbO_4$ and $YTaO_4$. In the last compound, a phase with six coordinated Ta has been already found quenched from compressed natural mineral [53]. Therefore, it appears to be quite likely that the seven compounds here discussed will end up having at extreme pressure structures with the pentavalent cation coordinated by six O atoms. Another conclusion that can be extracted from Fig.11 is that, according with the structural sequence of $YMoO_4$, $YVO_4$, $YAsO_4$, and $YPO_4$, the stabilization of the scheelite structure occurs at higher pressure as the size of the pentavalent atom is reduced. The only compound that only partially matches with this trend is $YCrO_4$ in which apparently magnetic interactions could reduce the zircon-scheelite transition pressure [18].

Let us now discuss on the compressibility of the studied compounds. In all of them the compressibility of both zircon and scheelite phases is non isotropic. In zircon the *a*-axis is the most compressible one. The opposite happens in scheelite. The difference in the anisotropic of zircon and scheelite is related with the different ordering of $YO_8$ dodecahedra and $XO_4$ tetrahedra. The zircon structure can be considered as a chain of alternating edge-sharing $XO_4$ tetrahedra and $YO_8$ dodecahedra extending parallel to the *c*-axis, with the chain joined along the *a*-axis by edge-sharing $YO_8$ dodecahedra (see Fig. 1). Upon compression in both structures the $XO_4$ tetrahedra are less compressible than the $YO_8$ bidisphenoids. In the zircon structure, this makes the *c*-axis less compressible than the *a*-axis as observed. As a consequence of the symmetry changes between the zircon and the scheelite structures, a rearrangement of the $XO_4$ and $YO_8$ units takes place [54], see Fig. 1. In particular, in the scheelite structure, the $XO_4$ tetrahedra are aligned along the *a*-axis, whereas along the *c*-axis the $YO_8$ dodecahedra are intercalated between the $XO_4$ tetrahedra. Therefore, in this structure the *a*-axis is the less compressible axis as found in experiments. These arguments also explain the anisotropic thermal expansion of compounds like $YAsO_4$ [55].

The fact that upon compression most of the volume contraction of the crystal is coming from the reduction of Y-O bonds can be used to correlate this distance with the bulk modulus of the crystals. Results summarized in Table III support this hypothesis. Based upon it, a phenomenological relation was proposed to estimate the bulk modulus of zircon and scheelite-structured oxides: $B_0 = 610\ Z_i / d^3$, where in our case $Z_i$ is the cationic formal charge of yttrium, $d$ is the mean Y – O distance at ambient pressure (in Å), and $B_0$ is given in GPa [37]. The same relationship applies also well to other ternary oxides like wolframites [56]. In Table III we summarize the bulk modulus calculated and experimentally determined for different compounds. The agreement among theory, experiments, and the empirical model is good. In order to extend the study on compressibility we theoretically determined the EOS of β-fergusonite $YNbO_4$: $V_0$ = 307.56 $Å^3$, $B_0$ = 132.63 GPa, $B_0'$ = 4.54, and $B_0''$ = -0.299 $10^{-3}$ $GPa^{-1}$, $YTaO_4$: $V_0$ = 303.91 $Å^3$, $B_0$ = 147.32 GPa, $B_0'$ = 3.98, and $B_0''$ = -0.213 $10^{-3}$ $GPa^{-1}$; and $YMoO_4$: $V_0$ = 307.84 $Å^3$, $B_0$ = 118.32 GPa, $B_0'$ = 6.34, and $B_0''$ = -0.701 $10^{-3}$ $GPa^{-1}$. The calculated bulk modulus are match reasonable well by the values obtained using the empirical relationship (see Table III), which indicates that also in these three compounds compressibility is mostly accounted by the reduction of Y – O bonds.

**B. Raman-active phonons in the zircon structure**

As we explained above, it is known that there are 12 Raman-active modes for the zircon structure. Seven of these ($2A_{1g}$ + $2B_{1g}$ + $B_{2g}$ + $2_{Eg}$) are internal modes [57] corresponding, in the case of the $RXO_4$ compounds, to vibrations of oxygen atoms in the $(XO_4)^{3-}$ units, four ($2B_{1g}$ + $2E_g$) are external translational due to translations of the $(XO_4)^{3-}$ and $R^{3+}$ ions, and one ($Eg$) is rotational of whole $(XO_4)^{3-}$ units. In the case of $YAsO_4$, we have measured only six internal modes. The missing $E_g$ mode is a typical feature of zircon-type $YXO_4$ compounds. Indeed it has been only measured in $YPO_4$ [58]. The internal $B_{2g}$, one external $B_{1g}$, and the rotational $E_g$ modes are very close in wavelength in $YPO_4$ and become nearly identical in $YVO_4$ [59], see Table V. Probably, the same phenomenon occurs in $YAsO_4$ and $YCrO_4$ causing that only one of the three phonons has been measured. The different modes of $YAsO_4$ and $YCrO_4$ are compared with those of $YVO_4$ and $YPO_4$ in Table V and Fig. 12. There it can be seen that the frequencies of the internal and external modes are found to decrease in going from $YPO_4$ to $YAsO_4$ having the higher frequency the phonon of those compounds where the X atom has a smaller atomic mass. The three high-frequency internal modes ($E_g$, $B_{1g}$, and $A_{1g}$) are the only three modes not following exactly this systematic behavior.

Apparently, the frequency change could be related to changes in X-O bond lengths as proposed by Podor [60] as observed in rare-earth arsenates, vanadates, and phosphates [61]. Another interesting feature to discuss is the remarkable differences of the symmetric stretching mode ($\nu_1$), the $A_{1g}$ mode with higher wavelength, and the anti-symmetric stretching modes ($\nu_3$), the $B_{1g}$ and $E_g$ mode with higher wavelength. For $YPO_4$ the following sequence holds: $\nu_3(B_{1g}) > \nu_3(E_g) > \nu_1(A_{1g})$. For $YCrO_4$ and $YVO_4$ the sequence is: $\nu_1(A_{1g}) > \nu_3(E_g) > \nu_3(B_{1g})$. And for $YAsO_4$ the sequence is: $\nu_1(A_{1g}) > \nu_3(B_{1g}) > \nu_3(E_g)$. The change in the order of these modes is related to the change in covalence of the X-O bond [62].

Regarding pressure evolution, it can be seen that they are similar in $YCrO_4$, $YVO_4$, $YAsO_4$, and $YPO_4$ (see Refs. 8, 10, 41 and Table II). In all the compounds the high-frequency internal modes have large pressure coefficients. It is also a typical picture of the zircon phase in the four compounds the softening of several modes, which is related to mechanical instabilities induced by pressure. Finally, another typical feature of the zircon phase is the large Grüneisen parameter of one low-frequency $E_g$ mode, which causes a phonon crossing in $YAsO_4$.

## VI. Concluding remarks

In this work we reported an experimental and theoretical study of the structural stability of $YAsO_4$ and $YCrO_4$ under compression. X-ray diffraction experiments together with calculations have allowed us to determine the occurrence of a phase transition from the zircon to the scheelite structure. In addition, total-energy calculations predict the following structural sequence for $YAsO_4$: zircon → scheelite → $SrUO_4$-type. The last transition is predicted at 32 GPa and sixfold arsenic coordination is attained for the $SrUO_4$-type structure. The whole sequence of transitions is discussed in comparison with $YVO_4$, $YPO_4$, $YNbO_4$, $YMoO_4$, and $YTaO_4$. The room-temperature equations of state of different compounds are also reported and discussed. Furthermore, a comparative discussion of lattice-dynamics properties is presented. Finally, the band-gap for $YAsO_4$ and $YCrO_4$ and band structure for $YAsO_4$ are also reported.

**Acknowledgments**

We acknowledge the financial support of the Spanish MCYT (Grants MAT2010-21270-C04-01/03 and CSD2007-00045). X-ray diffraction experiments were performed at HPCAT (Sector 16), Advanced Photon Source (APS), Argonne National Laboratory. HPCAT is supported by CIW, CDAC, UNLV and LLNL through funding from DOE-NNSA, DOE-BES and NSF. APS is supported by DOE-BES, under DE-AC02-06CH11357. The UNLV HPSEC was supported by the U.S. DOE, National Nuclear Security Administration, under DE-FC52-06NA26274. Supercomputer time has been provided by the Red Española de Supercomputación and the MALTA cluster.

## References

[1] R. J. Finch and J. M. Hanchar, Reviews in Mineralogy and Geochemistry **53**, 1 (2003); DOI: 10.2113/0530001.

[2] H. Nyman, B. G. Hyde, and S. Andersson, Acta Cryst. B **40**, 441 (1984).

[3] M. Lang, F.X. Zhang, J. Lian, C. Trautmann, R. Neumann, and R.C. Ewing, Earth Planet. Sci. Lett. **269**, 291 (2008) and references therein.

[4] A. A. Kaminskii M. Bettinelli, A. Speghini, H. Rhee, H. J. Eichler, and G. Mariotto, Laser Phys. Lett. **5**, 367 (2008) and references therein.

[5] D. Errandonea and F. J. Manjón, Prog. Mater. Sci. **53**, 711 (2008).

[6] E. Knittle and Q. Williams, Am. Mineral. **78**, 245 (1993).

[7] D. Errandonea, R. S. Kumar, L. Gracia, A. Beltrán, S. N. Achary, and A. K. Tyagi, Phys. Rev. B **80**, 094101 (2009).

[8] Y. W. Long, L. X. Yang, Y. Yu, F. Y. Li, R. C: Yu, and C. Q. Jin, Phys. Rev. B **75**, 104402 (2007).

[9] E. Climent Pascual, J. M. Gallardo Amores, R. Sáez Puche, M. Castro, N. Taira, J. Romero de Paz, and L. C. Chapon, Phys. Rev. B **81**, 174419 (2010).

[10] F. X. Zhang, J. W. Wang, M. Lang, J. M. Zhang, R. C. Ewing, and L. A. Boatner, Phys. Rev. B **80**, 184114 (2009).

[11] R. Lacomba-Perales, D. Errandonea, Y. Meng, and M. Bettinelli, Phys. Rev. B **81**, 064113 (2010).

[12] D. Errandonea, R. Lacomba-Perales, J. Ruiz-Fuertes, A. Segura, S. N. Achary, and A. K. Tyagi, Phys. Rev. B **79**, 184104 (2009).

[13] X. Wang, I. Loa, K. Syassen, M. Hanfland, and B. Ferrand, Phys. Rev. B **70**, 064109 (2004).

[14] M. Strada and G. Schwendimann, Gazzeta Chimica Italiana **64**, 662 (1934).

[15] B. A. Goldin, N. P. Yushkin, and M. J. Fishman, Zapinski Vsesoyuznogo Mineralogicheskogo Obshchstva **96**, 699 (1967).

[16] L. Gracia, A. Beltran, and D. Errandonea, Phys. Rev. B **80**, 094105 (2009).

[17] A. K. Pradhan, R. N. P. Choudhary, and B. M. Wanklyn, Phys. Status Sol. B **139**, 337 (1987).

[18] Y.W. Long, L.X. Wang, Y. Yu, F.Y. Li, R.C. Yu, S. Ding, Y.L. Liu, and C.Q. Jin, Phys. Rev B **74**, 054110 (2006).

[19] S. Klotz, J. C. Chervin, P. Munsch, and G. Le Marchand, J. Phys. D **42**, 075413 (2009).

[20] D. Errandonea, Y. Meng, M. Somayazulu, and D. Häusermann, Physica B **355**, 116 (2005).

[21] H. K. Mao, J. Xu, and P. M. Bell, J. Geophys. Res. **91**, 4673 (1986).

[22] A. P. Hammersley, S. O. Svensson, M. Hanfland, A. N. Fitch, and D. Häusermann, High Press. Res. **14**, 235 (1996).

[23] A. C. Larson and R. B. Von Dreele, LANL Report No. 86–748 (2000).

[24] W. Kraus and G. Nolze, J. Appl. Crystallogr. **29**, 301 (1996).

[25] R. Lacomba-Perales, J. Ruiz-Fuertes, D. Errandonea, D. Martinez-Garcia, and A. Segura, EPL **83**, 37002 (2008).

[26] G. Kresse and J. Hafner, Phys. Rev. B **47**, 558 (1993)

[27] G. Kresse and J. Furthmüller, Phys. Rev. B **54**, 11169 (1996).

[28] P.E. Blöchl, Phys. Rev. B **50**, 17953 (1994).

[29] J. P. Perdew, K. Burke, and M. Ernzerhof, Phys. Rev. Lett. **77**, 3865 (1996).

[30] H.J. Monkhorst and J.D. Pack, Phys. Rev. B **13**, 5188 (1976).

[31] G. Kresse, J. Furthmüller, and J. Hafner, Europhys. Lett. **32**, 729 (1995).

[32] K. Parlinski, Computer Code PHONON; software available at http://wolf.ifj.edu.pl/phonon.

[33] D. Errandonea, Phys. Status Solidi B **242**, R125 (2005).

[34] D. Errandonea, R. S. Kumar, X. Ma, and C. Y. Tu, J. Solid State Chem. **181**, 355 (2008).

[35] F. Birch, J. Geophys. Res. **83**, 1257 (1978).

[36] H. Li, S. Zhang, S. Zhou, and X. Cao, Inorganic Chemistry **48**, 4542 (2009).

[37] D. Errandonea, J. Pellicer-Porres, F. J. Manjón, A. Segura, Ch. Ferrer-Roca, R. S. Kumar, O. Tschauner, P. Rodríguez-Hernández, J. López-Solano, S. Radescu, A. Mujica, A. Muñoz, and G. Aquilanti, Phys. Rev. B **72**, 174106 (2005).

[38] L. Li, W. Yu, and C. Jin, Phys. Rev. B **73**, 174115 (2006).

[39] P. Mogilevsky, E. Zaretsky, T. Pathasarathy, and F. Meisenkothen, Phys. Chem. Miner. **33**, 691 (2006).

[40] J. Zhang, J. Wang, H. Zhang, C. Wang, H. Cong, and L. Deng, J. Appl. Phys. **102**, 023516 (2007).

[41]. F.J. Manjon, P. Rodriguez-Hernandez, A. Muñoz, A.H. Romero, D. Errandonea, and K. Syassen, Phys. Rev. B **81**, 075202 (2010).

[42] J. Ruiz-Fuertes, S. López-Moreno, D. Errandonea, J. Pellicer-Porres, R. Lacomba-Perales, A. Segura, P. Rodríguez-Hernández, A. Muñoz, A. H. Romero, and J. González, J. Appl. Phys. **107**, 083506 (2010).

[43] M. Dion, M. Ganne, and M. Tournoux, Mater. Res. Bull. **16**, 1429 (1981).

[44] T. L. Cremers, P. G. Eller, and E. M. Larson, Acta Crystallogr., Sect. C: Cryst. Struct. Commun. **42**, 1684 (1986).

[45] J. López-Solano, P. Rodríguez-Hernández, A. Muñoz, O. Gomis, D. Santamaría-Perez, D. Errandonea, F. J. Manjón, R. S. Kumar, E. Stavrou, and C. Raptis, Phys. Rev. B **81**, 144126 (2010).

[46] D. Errandonea, D. Martínez-García, A. Segura, J. Haines, E. Machado-Charry, E. Canadell, J. C. Chervin, and A. Chevy, Phys. Rev. B **77**, 045208 (2008).

[47] M. R. Dolgos, A. M. Paraskos, M. W. Stolzfus, S. C. Yarnell, and P. M. Woodward, J. Solid State Chem. **182**, 1964 (2009).

[48] N.R.J. Poolton, A.J.J. Boss, G.O. Jones, and P. Dorenbos, J. Phys. Cond. Matter **22**, 185403 (2010).

[49] K.S. Sohn, I.W. Zeon, H. Chang, S.K. Lee, and H.D. Park, Chem. Mater. **14**, 2140 (2002).

[50] N. J. Stedman, A. K. Cheetham, and P. D. Battle, J. Mater. Chem. **4**, 1457 (1994).

[51] W. M. Kriven, P. Sarin, and L. F. Siah, Solid-Solid Phase Transf. in Inorganic Materials **2**, 1015 (2005).

[52] D. Errandonea and F. J. Manjon, Materials Research Bulletin **44**, 807 (2009).

[53] H. Hori, T. Kobayashi, R. Miyawaki, S. Matsubara, K. Yokoyama, and M. Shimizu, J. Mineral. and Petrol. Sciences **101**, 170 (2006).

[54] M. Florez, J. Contreras-Garcia, J. M. Recio, and M. Marques, Phys. Rev. B **79**, 104101 (2009).

[55] H. C. Schopper, W. Urban, and H. Ebel, Solid State Commun. **11**, 955 (1972).

[56] J. Ruiz-Fuertes, D. Errandonea, R. Lacomba-Perales, A. Segura, J. González, F. Rodríguez, F. J. Manjón, S. Ray, P. Rodríguez-Hernández, A. Muñoz, Zh. Zhu, and C. Y. Tu, Phys. Rev. B **81**, 224115 (2010).

[57] G.M. Begum, G.W. Beall, L.A. Boatner and W.J. Gregor, J. Raman Spectrosc. **11**, 273 (1981).

[58] M.E. Poloznikova and V.V. Fomichev, Russian Chemical Reviews **63**, 399 (1994).

[59] R.J. Elliot, R.T. Harley, W. Hayes, and S.R.P. Smith, Proc. R. Soc. London A **328**, 217 (1972).

[60] R. Podor, Eur. J. Mineral. **7**, 1353 (1995).

[61] G. Barros, C.C: Santos, A.P. Ayala, I. Guedes, L.A. Boatner, and C.K. Loong, J. Raman Spectrosc. **41**, 694 (2010).

[62] J.A. Tosell, J. Am. Chem. Soc. **97**, 4840 (1975).

**Figure Captions**

**Figure 1:** Schematic view of the zircon (left), scheelite (center) and $SrUO_4$-type (right) structures. Large circles: trivalent metal (e.g. Y), middle-size circles: X atom (e.g. As, Cr, P, V). Small circles: oxygen atoms. The different polyhedral are shown as well as the unit cell.

**Figure 2:** Selection of diffraction patterns measured in $YAsO_4$ at different pressures. The pattern denoted with (r) was taken after pressure release. The reflections of the zircon and scheelite structures are shown by ticks for the patterns collected at ambient pressure. The crosses (+) and asterisks (*) indicate the appearance of peaks of the high-pressure phase and vestiges of the peaks of the low-pressure phase.

**Figure 3:** Pressure evolution of the unit-cell parameters of the zircon-type and scheelite-type phases of $YAsO_4$ and $YCrO_4$. To facilitate the comparison, for the HP phase we plotted $c/2$ instead of $c$. Symbols: experiments. Solid lines: linear fit. The vertical dotted lines indicate the onset of the phase transition and the pressure range of phase coexistence. Data for pressures < 8 GPa (4.2 GPa) for the scheelite phase in $YAsO_4$ ($YCrO_4$) were obtained on decompression.

**Figure 4:** Pressure-volume relation in $YAsO_4$ and $YCrO_4$. Symbols: experiments. Lines: EOS fit. Data for pressures < 8 GPa (4.2 GPa) for the scheelite phase in $YAsO_4$ ($YCrO_4$) were obtained on decompression.

**Figure 5:** Theoretical energy as a function of volume curves for the structures zircon (filled circles), scheelite (filled squares), $SrUO_4$-type (filled diamonds), monazite (filled triangles), iwashiroite (empty circles), $BaWO_4$-II-type (empty squares), $LaTaO_4$-type (empty diamonds), $BaMnF_4$-type (crosses), and $CuTeO_4$-type (empty triangles). Energy and volume are written per formula unit.

**Figure 6:** Theoretical (a) volume as a function of pressure and (b) enthalpy difference as a function of pressure curves for the most stable phases of $YAsO_4$ found in the calculations. Vertical dashed lines mark the theoretical transition pressures. In (b) at each pressure the enthalpy is measured with respect to the enthalpy of the zircon structure. Volume and enthalpy are written per formula unit.

**Figure 7:** Calculated pressure evolution of Raman modes for $YAsO_4$.

**Figure 8:** Calculated pressure evolution of IR modes for $YAsO_4$.

**Figure 9:** Band structure of zircon-type $YAsO_4$.

**Figure 10:** (color online) Atomic partial densities of states of zircon-type $YAsO_4$.

**Figure 11:** Structural sequence observed in different zircon-type oxides and related oxides. Data for $YVO_4$ and $YPO_4$ taken from Refs. [41] and [10]. The higher coordination phases box represent either the SrUO-type structure or any of the high-coordination phases found in related oxides; e.g. orthorhombic *Cmca* [5]

**Figure 12:** Schematic representation of Raman-active frequencies of $YAsO_4$, $YVO_4$ [41], $YCrO_4$, and $YPO_4$ [58]. In the abscissa the compounds are ordered in decreasing order according with the atomic number of the X atom.

**Table I**: Structural parameters of different structures of $YCrO_4$ and $YAsO_4$.

Zircon $YCrO_4$ at ambient pressure: $I4_1/amd$, Z = 4, *a* = 7.089(5) Å and *c* = 6.281(3) Å

| | Site | *x* | *y* | *z* |
|---|---|---|---|---|
| Y | 4a | 0 | 0.75 | 0.125 |
| Cr | 4b | 0 | 0.25 | 0.375 |
| O | 16h | 0 | 0.4283 | 0.2119 |

Zircon $YAsO_4$ at ambient pressure: $I4_1/amd$, Z = 4, *a* = 7.070(6) Å and *c* = 6.214(4) Å

| | Site | *x* | *y* | *z* |
|---|---|---|---|---|
| Y | 4a | 0 | 0.75 | 0.125 |
| As | 4b | 0 | 0.25 | 0.375 |
| O | 16h | 0 | 0.4283 | 0.2119 |

Scheelite $YAsO_4$ at ambient pressure: $I4_1/a$, Z = 4, *a* = 4.985(6) Å and *c* = 11.166(9) Å

| | Site | *x* | *y* | *z* |
|---|---|---|---|---|
| Y | 4b | 0 | 0.25 | 0.625 |
| As | 4a | 0 | 0.25 | 0.125 |
| O | 16h | 0.2434 | 0.0972 | 0.0450 |

Scheelite $YCrO_4$ at ambient pressure: $I4_1/a$, Z = 4, *a* = 5.0004(6) Å and *c* = 11.259(9) Å

| | Site | *x* | *y* | *z* |
|---|---|---|---|---|
| Y | 4b | 0 | 0.25 | 0.625 |
| As | 4a | 0 | 0.25 | 0.125 |
| O | 16h | 0.2524 | 0.0995 | 0.0461 |

**Table II:** Raman modes of zircon-type $YAsO_4$ and $YCrO_4$ at 1 bar. [a] Natural crystal, this work. [b] Synthetic powder [17]. [c] Synthetic crystal [17]. [d] Synthetic powder, this work. [e] Synthetic powder [18]. We also include the calculated phonon frequencies, pressure coefficients, and Grüneisen parameters for $YAsO_4$.

| | $YAsO_4$ | | | | | | $YCrO_4$ | |
|---|---|---|---|---|---|---|---|---|
| Mode | This work | Literature | | Calculations | | | This Work | Literature |
| | $\omega$ (cm$^{-1}$)[a] | $\omega$ (cm$^{-1}$)[b] | $\omega$ (cm$^{-1}$)[c] | $\omega$ (cm$^{-1}$) | $\partial\omega/\partial P$ (cm$^{-1}$/GPa) | $\gamma$ | $\omega$ (cm$^{-1}$)[d] | $\omega$ (cm$^{-1}$)[e] |
| $E_g$ | | | | 120.22 | -0.76 | -0.90 | | |
| $B_{1g}$ | 130.4 | | 130 | 155.76 | 1.28 | 1.17 | 148.8 | 149 |
| $E_g$ | 174.7 | 177 | 175 | 170.99 | -0.29 | -0.24 | 163.2 | 162 |
| $B_{1g}$ | | | | 208.95 | -1.66 | -1.13 | | |
| $E_g$ | 233.4 | 234 | 234 | 237.68 | 5.98 | 3.57 | | |
| $B_{2g}$ | 253.9 | 255 | 255 | 243.98 | 2.38 | 1.38 | 257.5 | 257 |
| $A_{1g}$ | 393.8 | 395 | 393 | 379.64 | 0.89 | 0.33 | 361.4 | 362 |
| $E_g$ | | | | 403.72 | 0.89 | 0.31 | | |
| $B_{1g}$ | 483.4 | | 481 | 474.17 | 2.03 | 0.61 | 552.3 | 552 |
| $E_g$ | 835.1 | 835 | 830 | 780.52 | 6.45 | 1.17 | 837 | 838 |
| $B_{1g}$ | 881.5 | 880 | 878 | 830.33 | 6.63 | 1.13 | 776.8 | 778 |
| $A_{1g}$ | 888.8 | 888 | 882 | 839.97 | 6.71 | 1.13 | 863.1 | 863 |

**Table III:** Ambient pressure bulk modulus for different phases of $YPO_4$, $YAsO_4$, $YCrO_4$, $YVO_4$, $YNbO_4$, and $YTaO_4$.

| Compound | Structure | Experiments | Theory | Empirical model |
|---|---|---|---|---|
| $YPO_4$ | zircon | 132[a],149(2)[b], 186(5)[c] | 144.4[f], 165.5[c] | 143 |
| $YAsO_4$ | zircon | 142(2)[d] | 115.6[d], 137[f] | 139.5 |
| $YCrO_4$ | zircon | 136(3)[d] | 121.6[g] | 139 |
| $YVO_4$ | zircon | 130(3)[e] | 122[h], 123.3[i], 142.8[h], 150.4[i] | 138 |
| $YPO_4$ | scheelite | | 213.7[c] | |
| $YAsO_4$ | scheelite | 149[d] | 135.2[d] | 148 |
| $YCrO_4$ | scheelite | 151[d] | 141.2[g] | 142 |
| $YVO_4$ | scheelite | 138[e] | 144.4[i] – 159.7[i] | 160 |
| $YMoO_4$ | scheelite | | 118.3 | 139.5 |
| $YNbO_4$ | fergusonite | | 132.6 | 139 |
| $YTaO_4$ | fergusonite | | 147.3 | 140.6 |

[a] Ref. [39]. [b] Ref. [11]. [c] Ref. [10]. [d] This work. [e] Ref. [13]. [f] Ref. [36]. [g] Ref. [38]. [h] Ref. [40]. [i] Ref. [41].

**Table IV:** Calculated orthorhombic $YAsO_4$ at 34.5 GPa: *Pbcm*, Z = 4, *a* = 4.8934 Å, *b* = 6.8098 Å, and *c* = 6.8414 Å.

| | Site | *x* | *y* | *z* |
|---|---|---|---|---|
| Y | 4d | 0.47932 | 0.18652 | 0.25 |
| As | 4a | 0 | 0 | 0 |
| O | 8a | 0.31600 | 0.58547 | 0.57197 |
| O | 4c | 0.83849 | 0.75 | 0 |
| O | 4d | 0.87051 | 0.02670 | 0.25 |

**Table V:** Calculated IR (up section) and silent modes (bottom section) of zircon-type $YAsO_4$ at 1 bar. The pressure coefficients and Grüneisen parameters are also included. The results are compared with experimental data [17]. [a] Synthetic powder. [b] Synthetic crystal.

| Mode | Literature | | Calculations | | |
|---|---|---|---|---|---|
| | ω ($cm^{-1}$)[a] | ω ($cm^{-1}$)[b] | ω ($cm^{-1}$) | ∂ω/∂P ($cm^{-1}$/GPa) | γ |
| $E_u$ | 210 | 210 | 191.94 | 2.86 | 2.12 |
| $A_{2u}$ | 244 | 249 | 254.12 | 2.48 | 1.39 |
| $E_u$ | 278 | 270 | 287.74 | 1.17 | 0.58 |
| $E_u$ | 352 | 353 | 313.25 | 1.09 | 0.49 |
| $A_{2u}$ | 486 | 482 | 426.44 | 1.43 | 0.48 |
| $E_u$ | 812 | 810 | 777.78 | 6.71 | 1.22 |
| $A_{2u}$ | 830 | 828 | 808.77 | 6.41 | 1.12 |
| $B_{1u}$ | | | 98.51 | -5.18 | -7.47 |
| $A_{2g}$ | | | 188.96 | -1.47 | -1.10 |
| $A_{1u}$ | | | 312.89 | 0.46 | 0.21 |
| $B_{2u}$ | | | 461.66 | 2.10 | 0.65 |
| $B_{2u}$ | | | 826.24 | 8.09 | 1.39 |

**Table VI:** Wavenumbers (in $cm^{-1}$) of the Raman-active modes at ambient pressure of the $YXO_4$ (Y = As, Cr, P, V) compounds. Data on $YPO_4$ and $YVO_4$ were taken from Refs. 41 and 58. Data on $YAsO_4$ and $YCrO_4$ are from this work.

| Mode | | $YAsO_4$ | $YVO_4$ | $YCrO_4$ | $YPO_4$ |
|---|---|---|---|---|---|
| External Translation | $E_g$ | | | | 157 |
| | $B_{1g}$ | 130.4 | 156.8 | 149 | 185 |
| | $E_g$ | 174.7 | 163.2 | 162 | 209 |
| | $B_{1g}$ | | 259.6 | | 315 |
| Rotation | $E_g$ | 233.4 | 260.5 | | 299 |
| Internal | $B_{2g}$ | 253.9 | 260 | 257 | 332 |
| | $A_{1g}$ | 393.8 | 378.4 | 362 | 484 |
| | $E_g$ | | | | 581 |
| | $B_{1g}$ | 483.4 | 489 | 552 | 660 |
| | $E_g$ | 835.1 | 838 | 818 | 1027 |
| | $B_{1g}$ | 881.5 | 816 | 778 | 1058 |
| | $A_{1g}$ | 888.8 | 891.1 | 863 | 1001 |

**Figure 1**

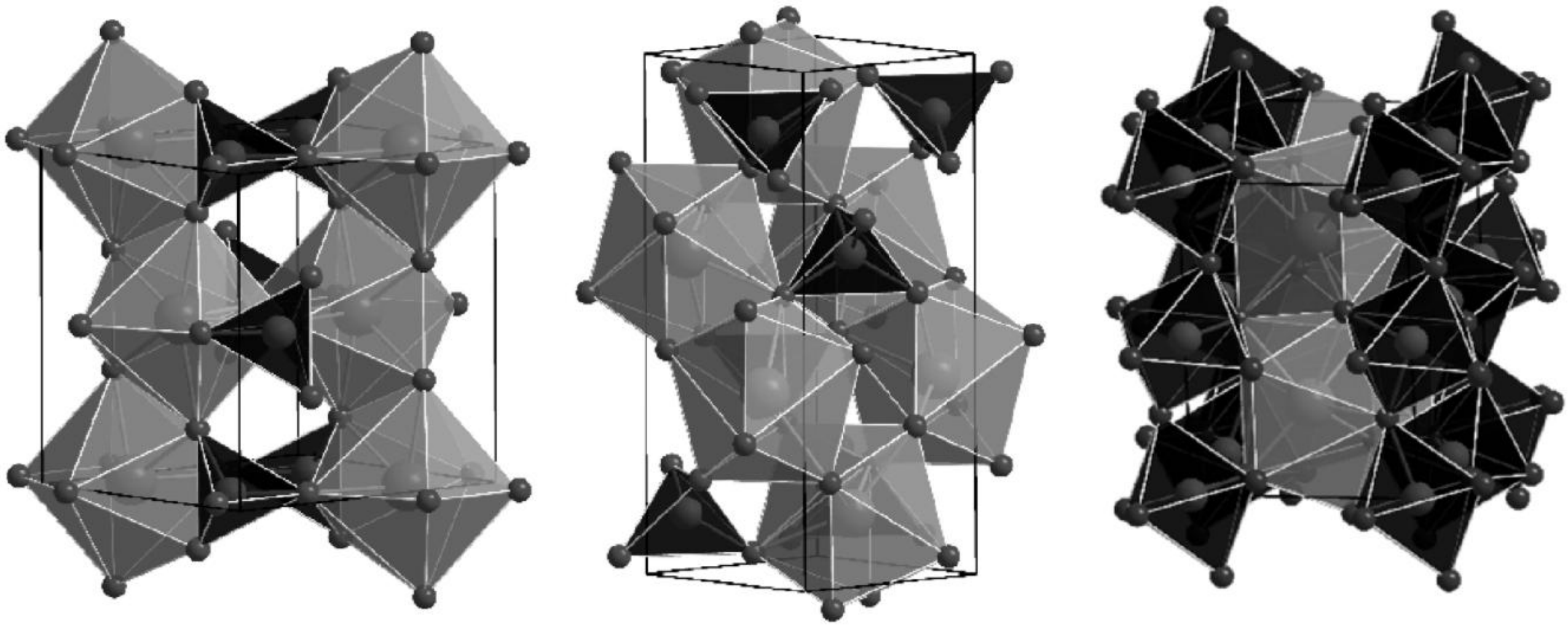

**Figure 2**

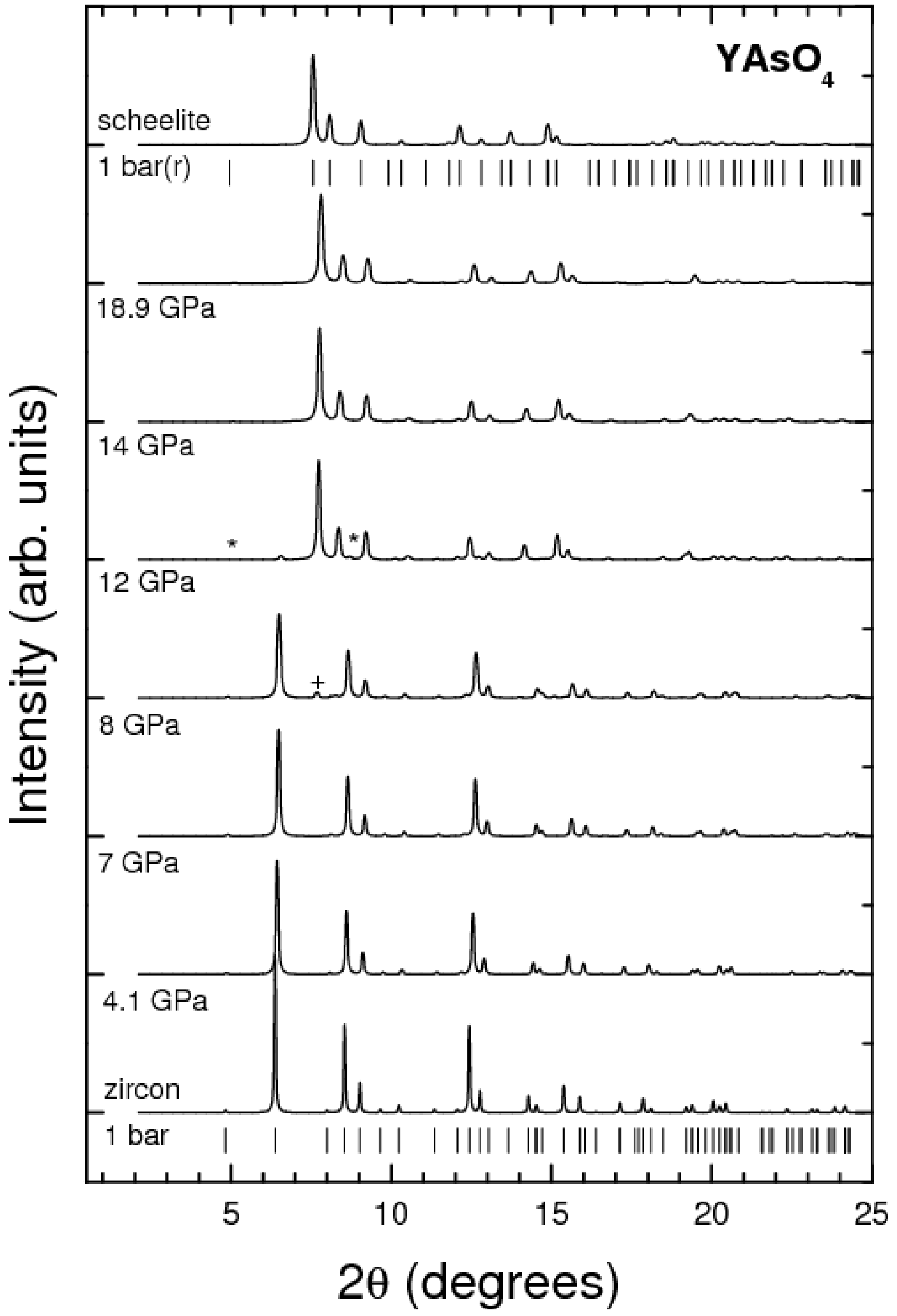

**Figure 3**

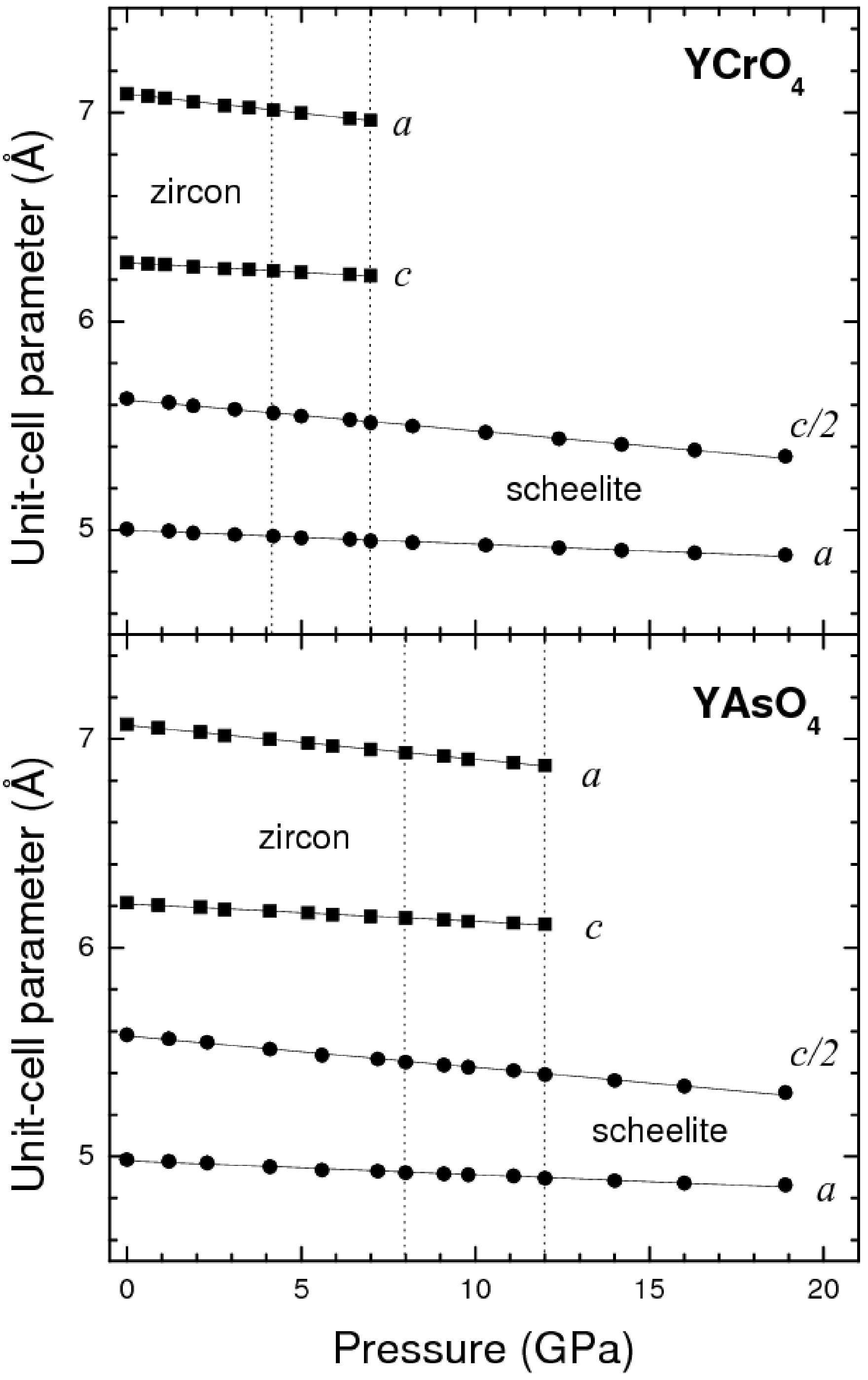

**Figure 4**

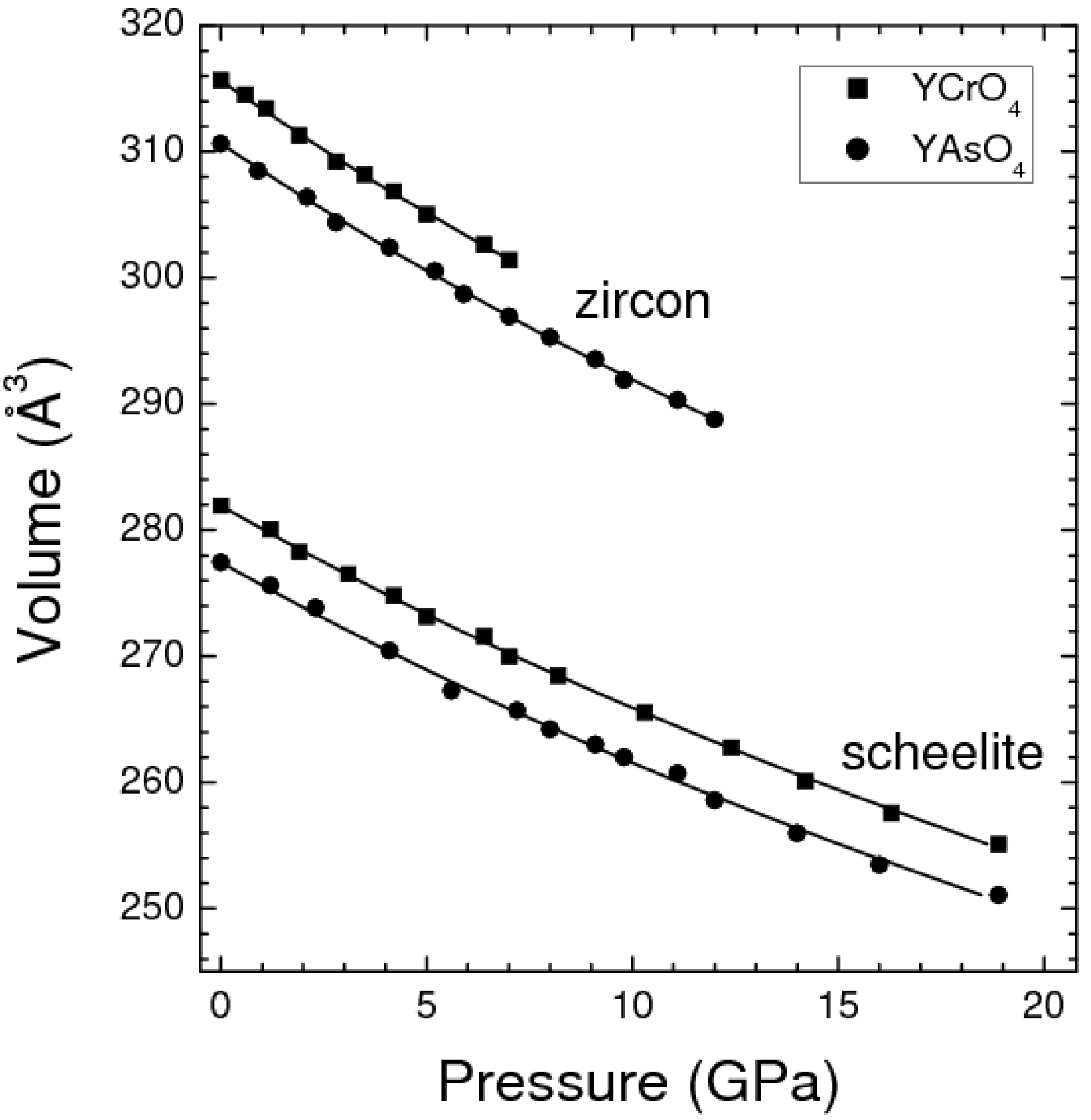

**Figure 5**

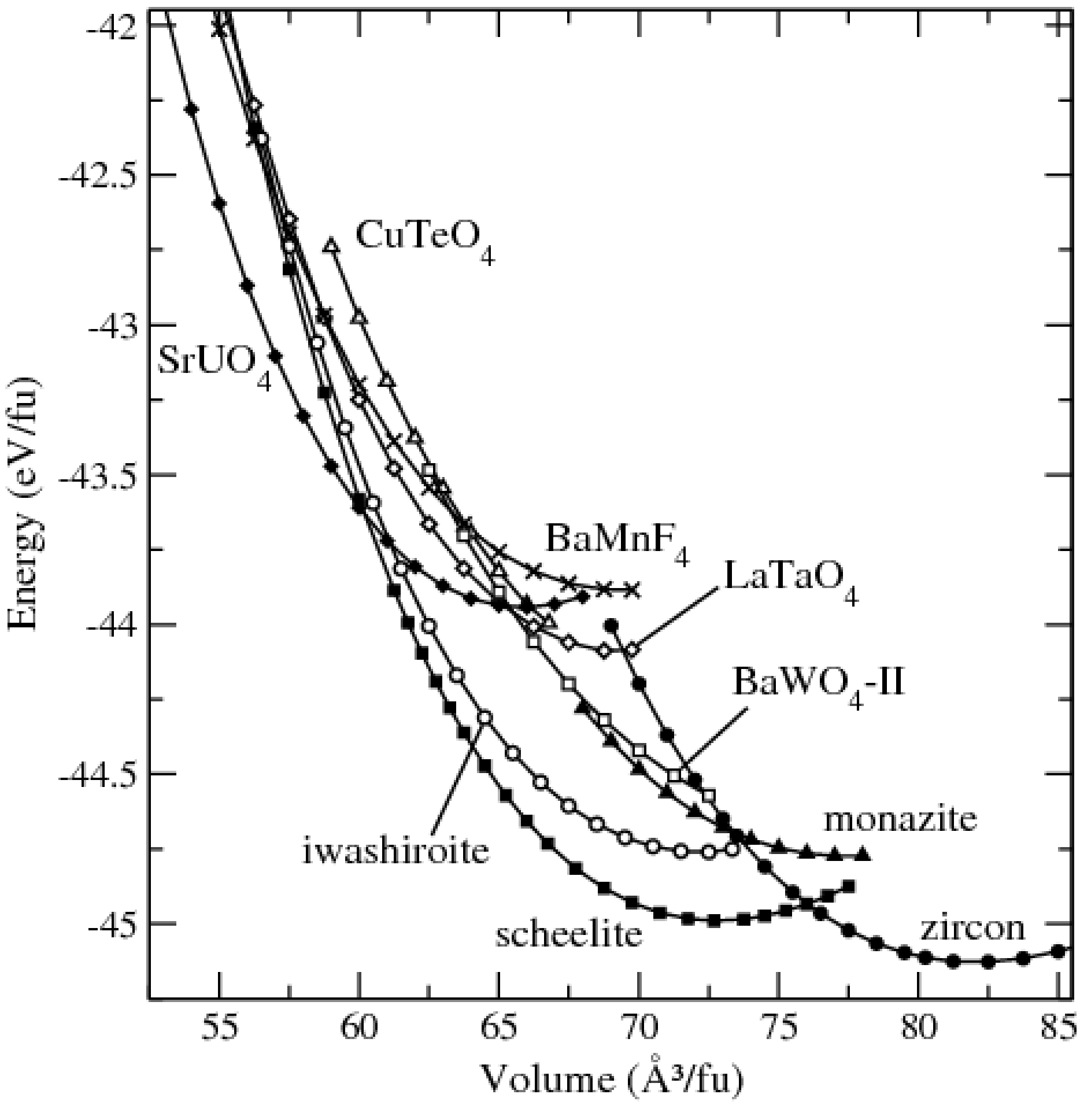

**Figure 6**

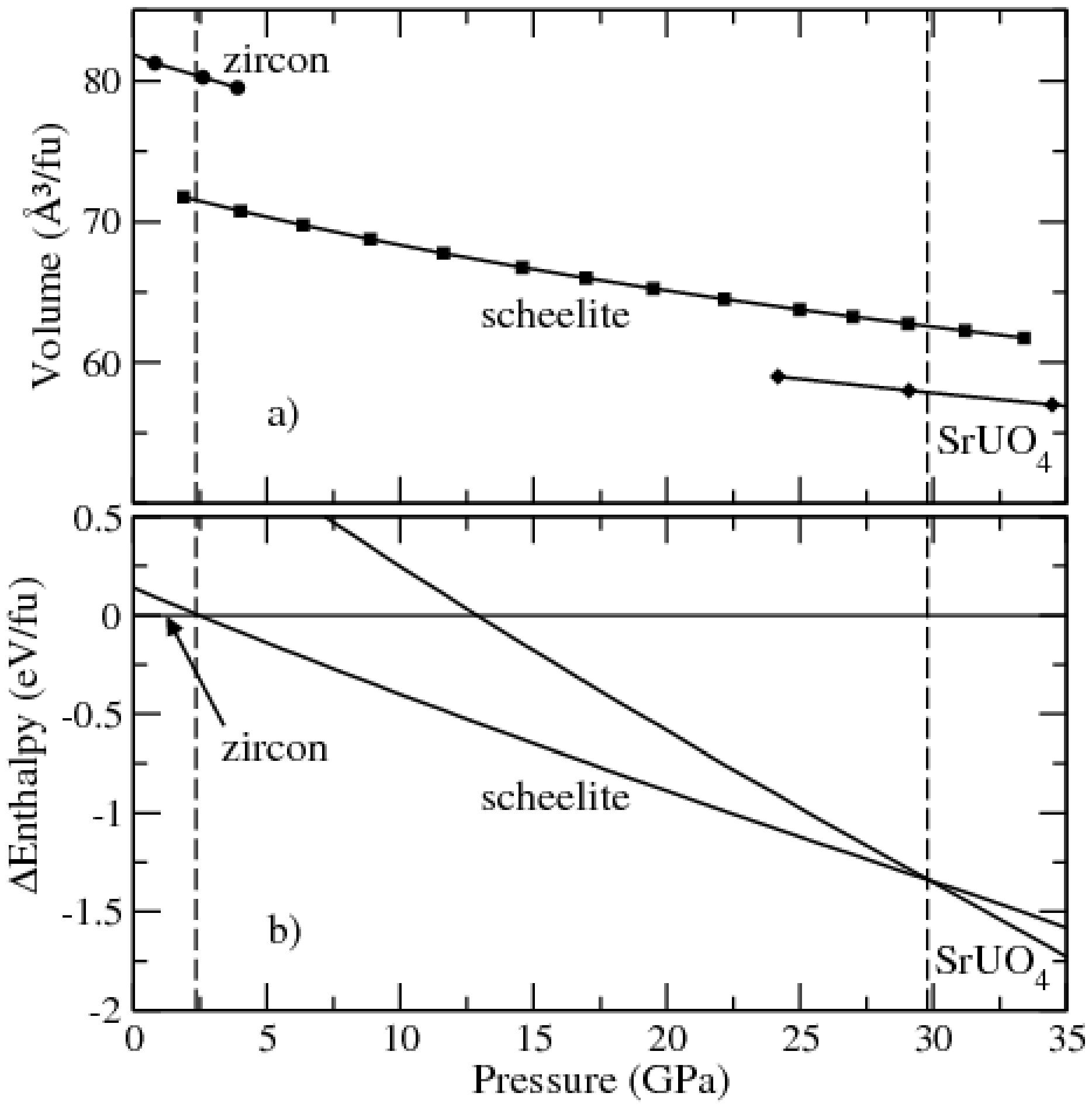

**Figure 7**

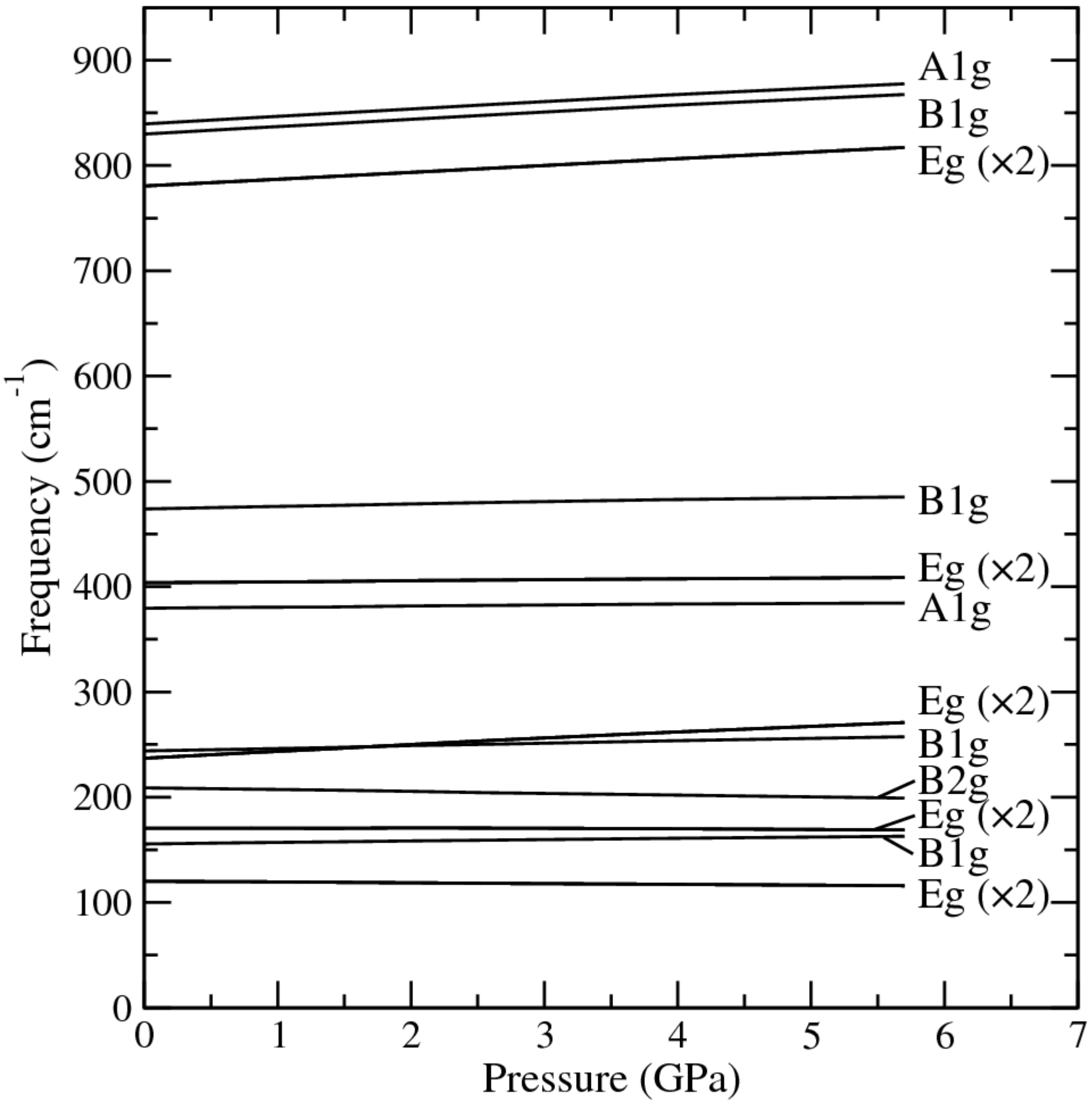

**Figure 8**

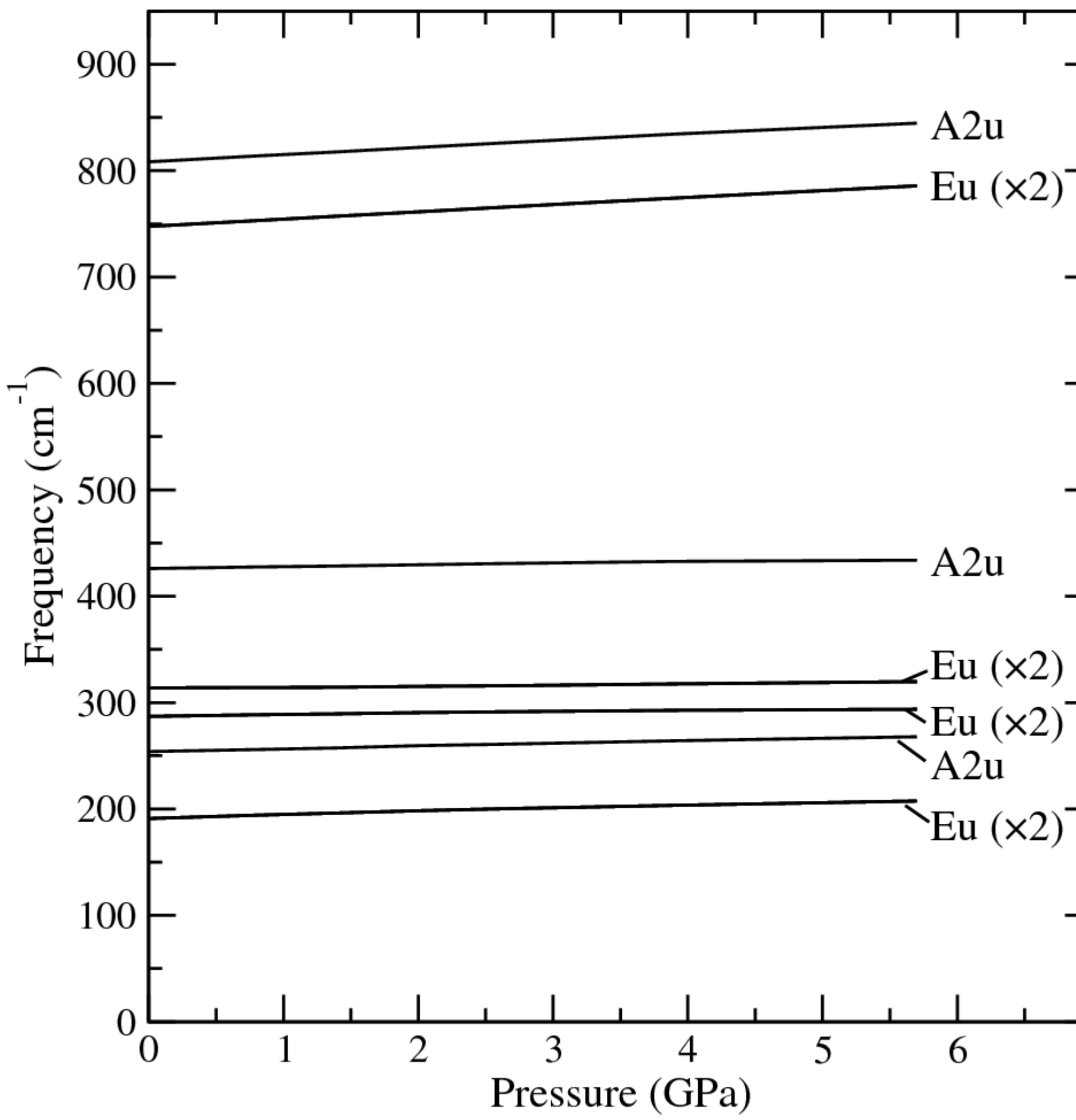

**Figure 9**

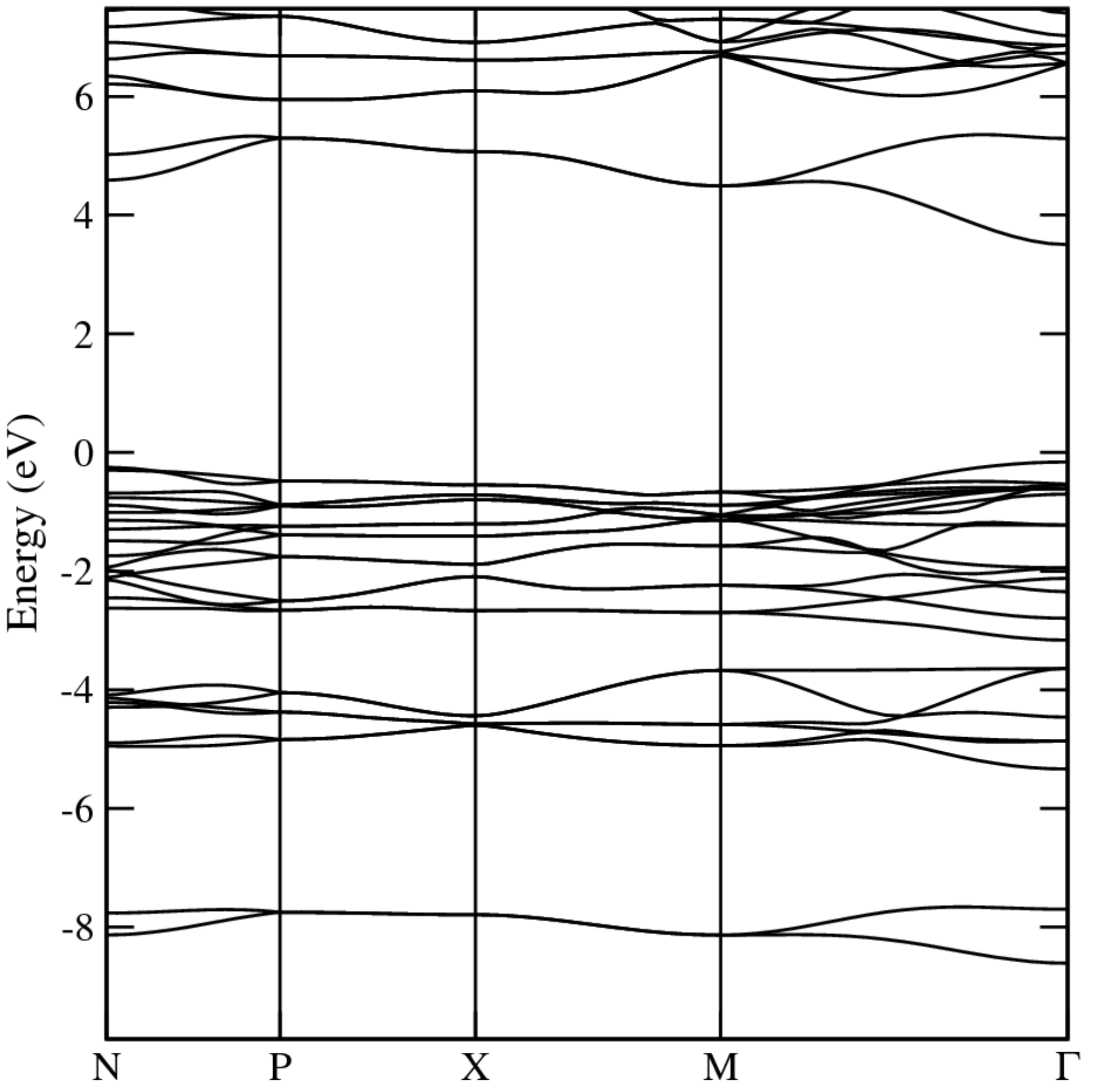

**Figure 10**

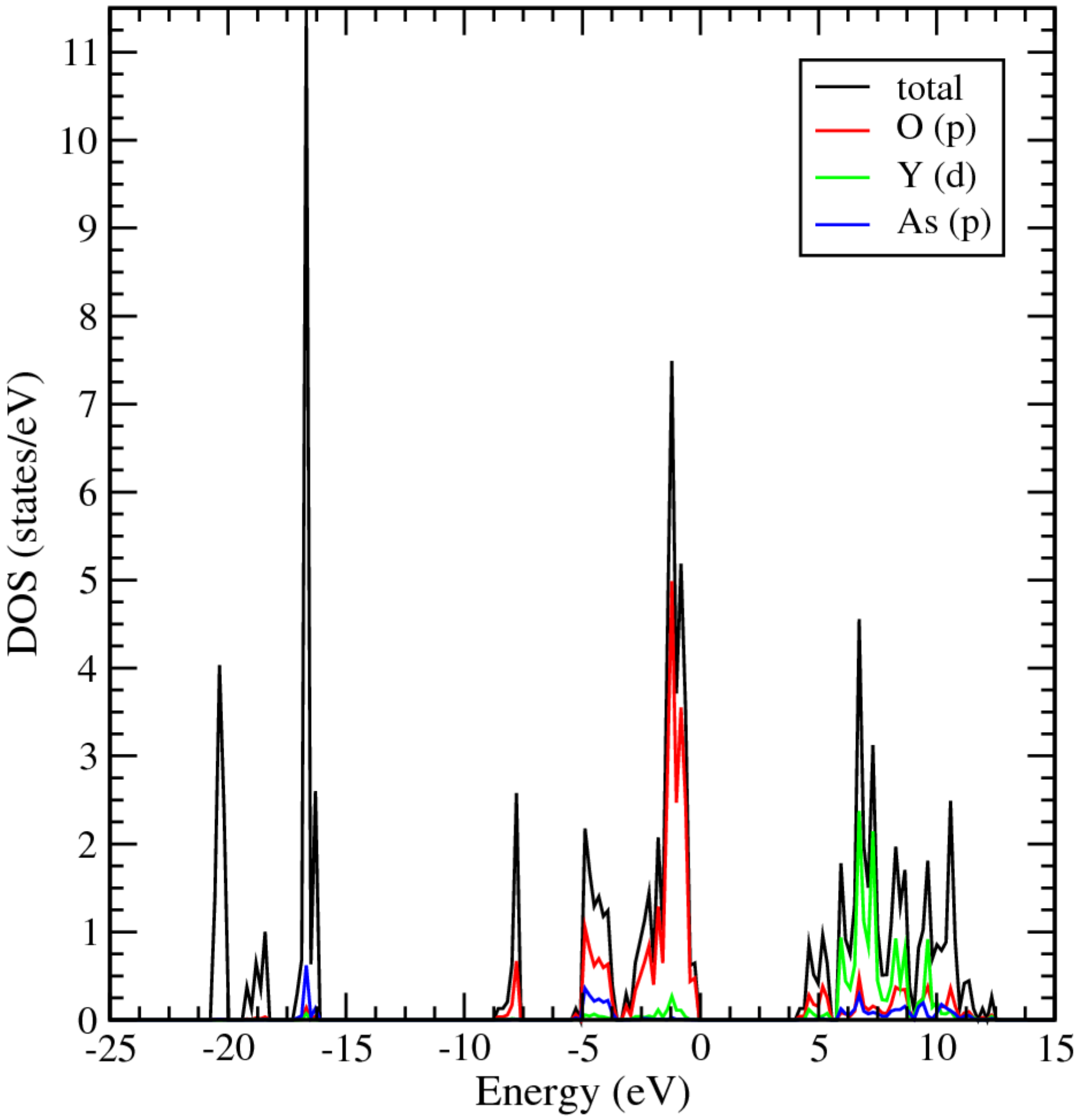

**Figure 11**

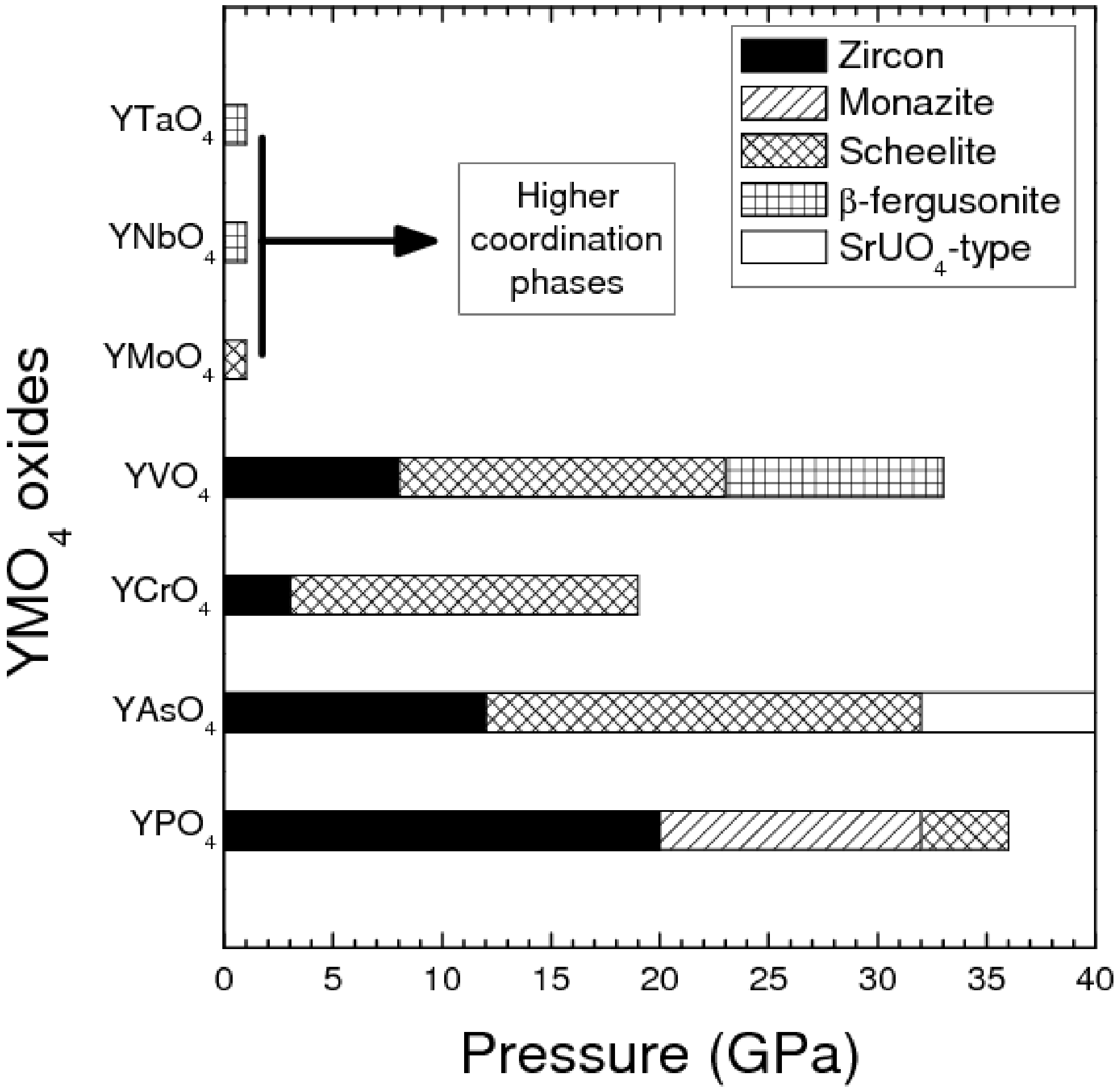
Zircon
Monazite
Scheelite
β-fergusonite
SrUO4-type
Higher coordination phases
YTaO4
YNbO4
YMoO4
YVO4
YCrO4
YAsO4
YPO4
YMO4 oxides
0 5 10 15 20 25 30 35 40
Pressure (GPa)

**Figure 12**

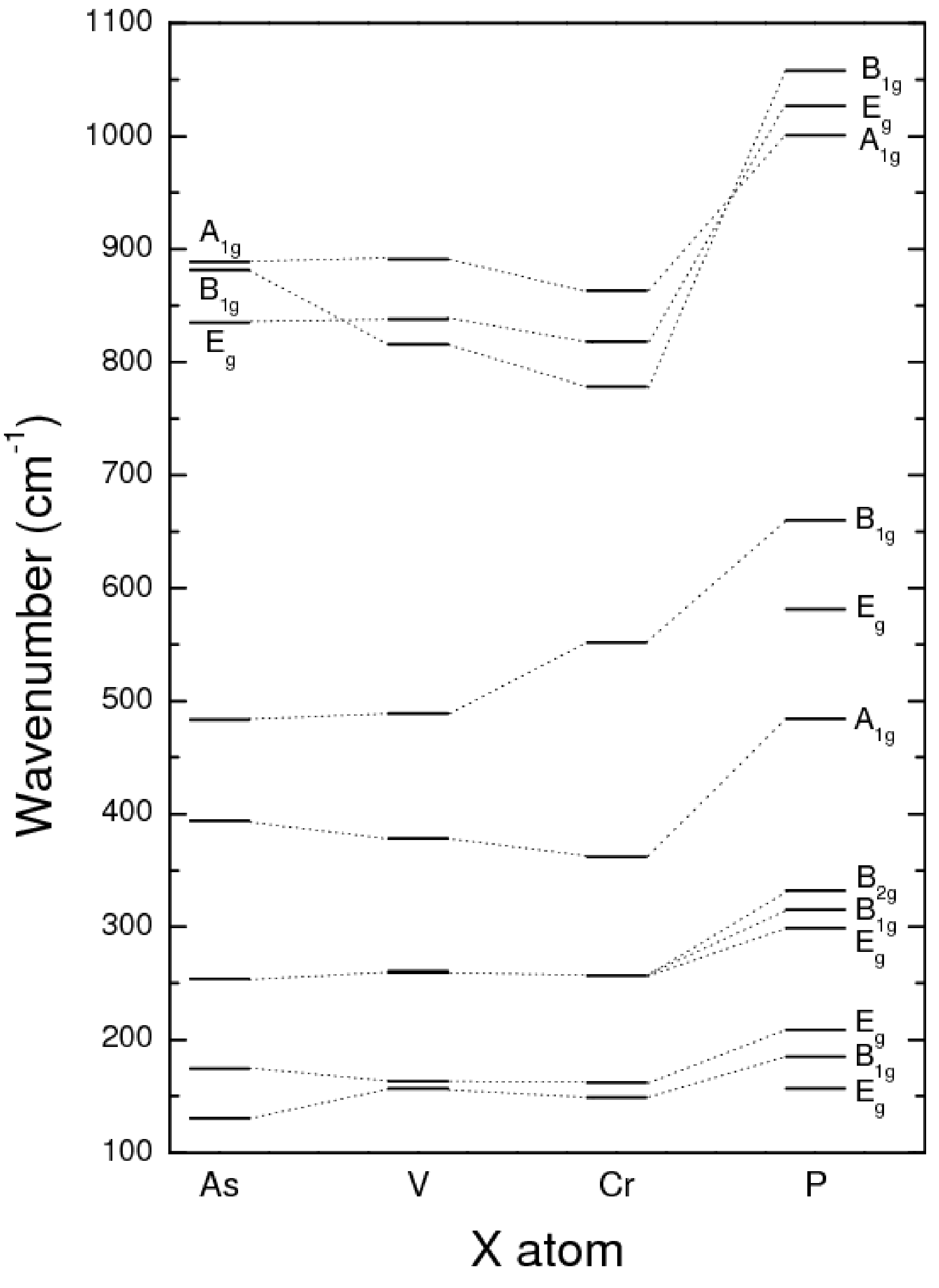